\documentclass{article}
\usepackage[a4paper,bindingoffset=0.5cm,left=1.5cm,right=1.5cm,top=2.5cm,bottom=2cm,footskip=.8cm]{geometry}
\usepackage[linesnumbered,ruled]{algorithm2e}
\usepackage{amssymb,eufrak,enumerate,makecell,float}
\usepackage{amsmath,mathrsfs,amsfonts,amsbsy, amsthm}
\usepackage{bbm}
\usepackage[table]{xcolor}
\usepackage{latexsym}
\usepackage{graphicx}
\usepackage{amsfonts}
\usepackage{epstopdf}
\usepackage{arydshln}
\usepackage{color}
\usepackage{caption}
\usepackage{subcaption}
\usepackage{hyperref}

\newtheorem{assumption}{Assumption}

\def\E{\mathbb{E}}

\newcommand{\Ints}{{\mathbb{Z}}}

\def\kron{\raisebox{1pt}{\ensuremath{\:\otimes\:}}}

\allowdisplaybreaks

\renewcommand{\tilde}{\widetilde}
\renewcommand{\hat}{\widehat}
\renewcommand{\bar}{\overline}
\renewcommand{\rho}{\varrho}

\newcommand{\REV}{{\omega}}
\newcommand{\COST}{{\theta}}

\DeclareMathOperator*{\argmax}{arg\,max}

\usepackage{booktabs}

\AtBeginDocument{%
  \providecommand\BibTeX{{%
    \normalfont B\kern-0.5em{\scshape i\kern-0.25em b}\kern-0.8em\TeX}}}

\begin{document}

\title{Optimising capacity allocation in networks of stochastic loss systems:\\ A functional-form approach}
\author{
Brendan Patch\\
School of Mathematics and Statistics\\The University of Melbourne\\
Australia\\
patch@unimelb.edu.au
\and
Mark S.\ Squillante\\
Mathematical Sciences\\IBM Research\\
United States of America\\
mss@us.ibm.com
\and
Peter M.\ van de Ven\\
Centrum Wiskunde \& Informatica (CWI)\\
The Netherlands
}


%

\maketitle

\begin{abstract}
{Motivated by a wide variety of applications, this paper introduces a general class of networks of stochastic loss systems
in which congestion renders lost revenue due to customers or jobs being permanently removed from the system.}
We
seek to balance the trade-off
between mitigating congestion by increasing service capacity and maintaining low costs for the service capacity provided.
{Given the lack of analytical results and the computational burden of simulation-based methods, we}
%
propose a hybrid functional-form approach for finding the optimal resource allocation
{in general networks of stochastic loss systems that combines} the speed of an analytical approach with the accuracy of simulation-based optimisation.
The key insight is
a core {iterative} algorithm that replaces
the computationally expensive gradient estimation in simulation optimisation with a closed-form analytical approximation that is calibrated using a
simple simulation run. 
{Extensive} computational experiments on complex
{networks}
show that our approach
renders near-optimal solutions
{with}
objective function values that are comparable to those obtained using stochastic approximation,
surrogate optimisation and Bayesian optimisation methods while requiring significantly less computational effort.
\end{abstract}

\section{Introduction}\label{sec:Intro}
Stochastic networks encountered in practice often involve losses due to various forms of congestion.
A key design decision for such networks concerns the allocation of (expensive) resources to limit losses while maintaining low costs.
For example, in cellular wireless networks,
the calls of moving users may be dropped if the receiving cell tower has insufficient capacity available for a successful handover (see, e.g.,~\cite{agrawal1996channel,kwan2010mobility,sidi1997new}).
{
Given the ubiquity of wireless networks, the allocation of resources in these networks plays a crucial role in many real-world applications experienced throughout daily life.
}
A similar situation may arise
in mobile cloud computing, where computational jobs are served at small ``cloudlets'' close to the user, rather than at a remote cloud computing facility (see, e.g.,~\cite{Pang2015,Jia2016,mao2017survey}).
While
the use of cloudlets reduces
delays~\cite{Jia2016}, each cloudlet has limited capacity and migrating jobs between cloudlets due to user mobility may lead to job losses and performance issues.
{
Given the growth of such cloudlet-based mobile computing, resource allocation in these environments plays a critical role in various real-world applications
including augmented reality, crowd-sourcing video processing, and facial recognition~\cite{Pang2015}.
}
These processes {based on wireless networks and cloudlet mobile computing} can be formulated as part of a general class of networks of stochastic loss systems (SLSs). 

Another example within the general class of networks of
SLSs
concerns supply chain networks designed for
{the manufacturing and}
maintaining
{of computer and network systems}
with uncertain lifetimes.
Any downtime of these
{systems can be}
expensive and disruptive, and should be minimised to the extent possible.
When {such} a
{computer or network system}
breaks down, the necessary spare parts are quickly dispatched from the nearest stock point.
If a spare part is unavailable at this location, losses occur and the parts must be sent from a more distant central warehouse, thus increasing downtime (see, e.g.,~\cite{Kranenburg2009,Rahimi-Ghahroodi2017,shen2003joint})
{and in turn making resource allocation crucially important.}
{An analogous}
problem arises in the context of emergency services, where ambulances and fire trucks are geographically dispersed to ensure a fast response to emergencies
(see, e.g.,~\cite{brotcorne2003ambulance,vandenBerg2017,van2016comparison}).
Requests for these vehicles (i.e., a nearby emergency) may be rejected because they are already in service,
resulting in
the risk of loss of life
and requiring a rerouting of the request to other vehicles/resources
with associated delays.
{The allocation of ambulance and fire fighting resources, possibly through computer-automated support, obviously plays a critically important role in emergency services.}
These  processes {based on supply chain and emergency service networks} can also be formulated as part of a general class of networks of
SLSs.

In this paper we consider
such a general class of networks of
SLSs,
with a particular focus on both of the above stochastic network model instances within the general class.
%
%
{Given the importance of resource allocation in this class of networks of
SLSs,
we observe that there exist essentially two categories of capacity management approaches in stochastic networks in general:}
analytical and simulation-based.
Analytical approaches typically approximate the complex stochastic network with a much simpler network and solve the corresponding optimisation problem.
These approximations may be based on product-form networks (see, e.g.,~\cite{baskett1975open,harrison1992brownian,kelly1991loss}),
independence assumptions among network components
(see, e.g.,~\cite{axsater1990modelling,Restrepo2009}),
aggregation of network components (see, e.g.,~\cite{sonmez2017analytical}), 
fluid-limit scalings (see, e.g.,~\cite{harrison2005method}),
and large-scale systems (see, e.g.,~ \cite{borst2004dimensioning,Hassin2015}).
These closed-form approximations can then be used to solve the capacity management problem, either analytically (see, e.g.,~\cite{bookKleinrock1964,wein1988capacity}),
numerically, or using a combination of both (see, e.g.,~\cite{Sanders2016}).

Simulation-based approaches typically evaluate and update capacity allocation using simulation of the stochastic network.
One canonical simulation-based optimisation approach is stochastic approximation (SA) (see, e.g.,~\cite{asmussen2007stochastic,henderson2006handbooks,kiefer1952stochastic,robbins1985stochastic} and Appendix~\ref{app:StochApprox}), where gradients are estimated using simulation in order to apply gradient descent in a stochastic setting.
Similar adaptations of Newton and quasi-Newton
methods, based on higher-order derivatives, are currently a very active area of research
(see, e.g.,~\cite{byrd2016stochastic}). 
Another well-known approach is Bayesian optimisation (BO) (see, e.g.,~\cite{bull2011convergence,mockus2012bayesian}), where a prior is placed over the unknown objective function which is then updated via noisy observations. 
Surrogate optimisation (SO) is yet another popular approach (see, e.g., \cite{gutmann2001radial,jones2001taxonomy,razavi2012review}),
which optimises over a response surface that is generated by radial basis function interpolation between function evaluations.

Both analytical and simulation approaches have limitations that prevent application to the complex stochastic loss networks of interest.
In analytical approaches, it is often impossible to find an approximation for the complex network that is sufficiently accurate and captures all relevant features;
as a result, any capacity management decisions based on the approximate model may not, in general, work well in the original model.
Simulation-based approaches achieve better accuracy, but can suffer from large computational costs which often make their application to high-dimensional complex
stochastic loss networks{, as well as their application at finer time scales,} infeasible.
These large computational costs may be driven by the need to estimate a (high-dimensional) gradient, which requires many simulation runs to
obtain estimates with sufficient accuracy.

\begin{figure}
  \includegraphics[width=0.85\textwidth]{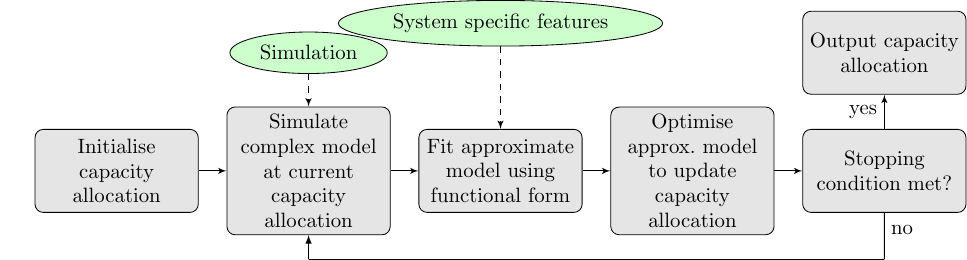}
  \caption{{Iterative} functional-form optimisation.}
 \label{fig:flow}
\end{figure}

A recent paper~\cite{Dieker2017}
introduced a general hybrid approach for a class of stochastic networks without losses that exploits its theoretical properties to significantly reduce the computational costs of simulation-based optimisation. 
{The proposed approach consists of two phases where}
simulation is used {in the first phase} to fit a closed-form functional-form approximation to the performance metric at the current resource allocation level. 
This
approximation is then used to analytically determine a new resource allocation, rather than using gradient descent, 
and the process is repeated iteratively.
A general version of the corresponding {first-phase} iterative process is illustrated in Figure~\ref{fig:flow}. 
It is shown in \cite{Dieker2017} that the resource allocation obtained {in the first phase} is close to optimal while only requiring a fraction of the computational effort of existing methods,
due to reducing the number of iterations and due to not having to estimate gradients from simulation in each iteration.
The approach is similar to response surface or surrogate optimisation methods (see, e.g., \cite[Section~3.2]{amaran2016simulation}),
but differs due to the significantly higher level of detail that is analytically garnered about system structure and incorporated into the functional-form approximation model. 
{Once the first-phase iterative process converges, the resulting near-optimal solution is then used as a starting point for the SA method as part of the proposed two-phase approach.}

The {first-phase} approximation approach of Dieker et al.~\cite{Dieker2017} does not carry over to the
general class of networks of
SLSs
studied in this paper, since the addition of losses fundamentally changes the dynamics and other characteristics of the stochastic network.
In particular, the objective changes from a function of the queue lengths in~\cite{Dieker2017} to a function of the probabilities of losses;
we are also interested in
discrete capacities for the number of servers at each station of the network rather than continuous per-station service rates.
Moreover, because the problem in~\cite{Dieker2017} considers each queue to have infinite buffer capacity and thus no jobs are lost,
the arrival rate into every station is independent of the capacity allocation subject to stability constraints.
Then under the {first-phase} functional-form approximation in \cite{Dieker2017}, the problem can be essentially decoupled into smaller sub-problems for each station.
Since losses remove jobs from the system and losses occur more frequently when there is reduced capacity,
we can no longer exploit in our setting the property that the effective long-term arrival rate into a station is independent of the capacity allocation of the upstream station(s).
This added level of interaction among the stations substantially increases the {problem} complexity.  

In this paper
our focus is on the first-phase approximation approach of Dieker et al.~\cite{Dieker2017} and
we propose a general iterative
approach
for optimising capacity allocation in networks of
SLSs
{that extends this first-phase}
approach in~\cite{Dieker2017}
{to address the important complexities of these networks}.
Our core {iterative} algorithm explicitly preserves the dependence among stations by replacing all appearances of a loss probability in the objective function with a unique functional-form
approximation---the capacity vector is then jointly optimised using numerical methods. 
The efficiency of our core algorithm is driven by the known relationships between losses at each station for each route, which are explicitly incorporated into the optimisation method. 
In order to encapsulate the full range of applications listed earlier, we detail our core algorithm for two model variants within the overarching general class of networks of
SLSs. 
The primary difference between the two models is the objective of the jobs in the network. 
For the first model (Model~I), each job aims to visit a subset of the stations and is lost as soon it encounters a station which is at capacity;
whereas for the second model (Model~II), each job aims to visit one station out of some subset of the stations and is only lost if all of the stations in the subset are at capacity.
The core {iterative} algorithm of our {first-phase} approximation approach
is not a generic simulation optimisation algorithm; it has been tailor-made specifically for the general class of networks of
SLSs
and should be used as such. 
At the same time, by
detailing two models within this general class we
demonstrate the general principles of how to use our {first-phase} approach within the class.

Our investigation into the performance of {the core iterative algorithm of} our
{first-phase} approximation approach
consists of extensive computational experiments on four example topologies of general networks of
SLSs:
Markovian tandem networks, criss-cross networks, ring networks, and a complex class of canonical networks.
We consider all four topologies in the context of both of the model variants introduced above.
Many of our examples feature renewal arrival processes and/or non-exponential service times and are thus represented by non-Markovian stochastic processes.
The first of our examples is (just) simple enough that a benchmark for the accuracy of our
core {iterative} algorithm
against true optimal solutions is possible for both model variants.
In addition to comparing the accuracy of our {core} algorithm with the exact solution for this simplest network topology example, 
we gauge the efficiency of our approach by benchmarking it against {the} SA, BO and SO algorithms {introduced earlier}. 
{We also use these benchmark algorithms to gauge the accuracy and efficiency of our core iterative algorithm for the three more complex example network toplogies},
SLSs
of interest,
with our focus on extending the first-phase approach in~\cite{Dieker2017} to support the general class of networks of
SLSs
of interest.

For the Markovian tandem network example, we observe that our {core} iterative algorithm {directly} achieves near-optimal solutions with very low computational effort.
By comparison,
{the other benchmark algorithms}
are occasionally slightly more accurate,
but even for this {relatively} simple model
{these benchmark algorithms by themselves}
already exhibit much
{higher computational costs.}
For the remaining {example topologies},
we observe that our {core} algorithm 
{makes tremendous strides within a few computationally inexpensive iterations to greatly improve the performance of the system from the initial starting point of random capacity allocation.}
We show that our {core} iterative algorithm compares {quite} favourably against SA both in terms of objective value found and in terms of computational efficiency.
We further provide evidence that our {core} algorithm is frequently faster than SO and usually much faster than BO. 
These massive benefits in {computational} efficiency
{are at worse realized with only}
a very minor reduction in {the} objective values found,
and indeed, for many of the scenarios we consider, our {core} algorithm outputs better solutions than the benchmark methods. 
Hence,
the solution of our core iterative algorithm can be either used at finer time scales to quickly obtain a high-quality solution or used at coarser time scales
to quickly provide a high-quality initial starting point for a preferred benchmark algorithm to efficiently obtain a refined solution, following the two-phase approach in~\cite{Dieker2017}.

In summary, our contributions are as follows: 
\begin{enumerate}
\item We introduce a new
general class of networks of
SLSs
with a wide range of applications across many areas, which {has been} previously ignored because of its intractability;
\item By extending the {first-phase} approach in \cite{Dieker2017}, we devise a
{novel core iterative algorithm of our general first-phase approximation approach}
that
{supports}
this new class of stochastic models;
\item We apply {our} {first-phase} approach to two model variants within the general class of networks of
SLSs,
and show that it performs remarkably well. 
\end{enumerate}

The remainder of this paper is organised as follows. 
Section~\ref{sec:Model} presents our general class of networks of
SLSs,
including two canonical instances of this class of stochastic network models together with their corresponding objective functions. 
Section~\ref{sec:FF4SNB} provides an overview of our {iterative} functional-form optimisation approach for {first-phase solutions to} our general class of networks of
SLSs.
Section~\ref{sec:Experiments} demonstrates that our core algorithm works well on realistic network examples under a variety of scenarios, outperforming several benchmark approaches. 
This is followed by concluding remarks in Section~\ref{sec:Conclude}. 
In addition, Appendix~\ref{app:MAM} shows how to compute the objective function for a small Markovian network instance, which is used as a benchmark in one of our computational experiments.%


\section{Mathematical models}\label{sec:Model}
We now define {a general class of} networks of
SLSs
and the corresponding optimisation problem formulation, followed by a small example for illustration {and elucidation}.
Consider a set of stations
{$\mathcal L = [L] := \{1,\dots,L\}$;}
e.g., each station could represent a base station in a cellular {wireless} network or a stock point in a {computer} spare-parts network. 
Station $l$ has $c_l$ servers that can each serve one customer at any given time; we refer to $c_l$ as the capacity of station $l$ and define
{$\boldsymbol c := {(c_l)}_{l\in [L]}$.}
Define
{$\mathcal R := [R]$}
to be
the set of customer classes. 
Customers of class $r \in \mathcal R$ arrive to the network according to some general exogenous arrival process with rate $\lambda_r$. 
No further assumptions are made on the arrival processes, allowing as examples renewal type with independent and identically distributed interarrival times
(see, e.g., \cite{Brown1975,Patch2015}) and Markovian arrival processes with non-identically distributed interarrival times (see, e.g., \cite{bookHe2014,Narayana1992}). 
{Associated with each class $r$ is}
a path comprising $N_r$ network stations defined by
{$\psi_r := {(l(r,i))}_{i\in [N_r]}$,}
{where} $l(r,i) \in \mathcal L$ {is the $i$-th station in the class-$r$ path}.
We assume no station appears more than once in any path $\psi_r$, i.e., $l(r,i) \neq l(r,j)$, $\forall~i\neq j$. 
A class-$r$ customer arrives to the network at station $l(r,1)\in\mathcal L$.

As pointed out in the introduction, two variants of this
{class of stochastic networks}
are to be distinguished. 
The key differences between {these} variants are the goal of {the} customers, the effect that this has on their behaviour in response to congestion, and the implications for the revenue stream of the network operator. 
Our general functional-form optimisation approach then exploits underlying theoretical properties of the objective function (i.e., existing domain knowledge) to augment information from simulation and speed up
the iteration process of finding a good approximation to the optimal solution.
The {two} variants are described as follows.
\begin{itemize}
\item In the first variant, referred to as Model~I, the objective of customers is to be served, without interruption, by all of the stations along their path. 
For this variant, when a customer arrives at a station to find no available server, it will immediately depart the network and we say the customer has been {\em lost}.
However, if a station to which a customer arrives has an available server, it is accepted for service and remains at the server of the station for a generally
distributed amount of time with finite expectation---we impose no other restrictions on the service distributions.
Upon completion of service at station $l(r,i)$, for
{$i\in [N_r-1]$,}
the customer transitions to station $l(r,i+1)$ and either receives service or is lost, depending on whether a server is available at the station.
The customer leaves the network after receiving service at station $l(r,N_r)$.
Customers who successfully traverse all the stations of their path $\psi_r$ pay an amount $\REV_r$ to the network operator. 
Customers who are lost while traversing the network do not pay the network operator. 
Model~I is suitable, for example, as a representation of a cellular wireless network where calls of moving users may be dropped if the receiving cell tower has insufficient available capacity for a successful handover. 
{A similar application is cloudlet-based mobile computing, where the jobs of mobile users may be lost and/or encounter performance issues if the limited capacity of cloudlets is not sufficient to properly handle
the migration of these jobs along the path of the user.}
\item In the second variant, referred to as Model~II, the objective of customers is to be served at one and only one of the stations along their path.
For this variant, when a customer arrives at a station to find no available server, it will immediately attempt service at the next station on its path. 
If a customer is denied service from all of the stations in its path, then it is {\em lost}. 
If a station to which a customer arrives has an available server, it is accepted for service and remains at the server of the station for a generally
distributed amount of time with finite expectation---we impose no other restrictions on the service distributions. 
Upon acceptance for service at station $l(r,i)$ a customer pays $\REV_{r,i}$ to the network operator. 
Typically we would expect that $\REV_{r,i+1} < \REV_{r,i}$, reflecting  a preference for stations that are earlier in the path (though not a requirement of our approach).
Upon completion of service at station $l(r,i)$ the customer departs the network. 
Model~II is suitable, for example, as a representation of a supply chain network where customers who cannot obtain a {spare part for a computer system} at their preferred depot will seek to obtain the part at other depots.
Another application is emergency services, where if all of the units for a particular district are occupied when requested, then units from nearby districts can be redirected at a higher cost (human and financial). 
\end{itemize}

{\bf Objective functions.}  Our
{study}
addresses the fundamental trade-off between maintaining low capacity costs and
{collecting}
high rewards
(by ensuring few losses).
{Consistent}
with the dynamics described above, the objective function takes on a different form depending on which variant of the model is being considered. 
For both variants, each server at station $l$ costs $\COST_l>0$ per unit time, {and we} define
{$\boldsymbol \COST := {(\COST_l)}_{l\in [L]}$.}
This means that
the network operator pays $\langle \boldsymbol c, \boldsymbol \COST \rangle$ per unit time, where $\langle\cdot\rangle$ denotes the usual Euclidean inner product. 
Let $p_{r,i}(\boldsymbol c)$ denote the fraction of class-$r$ jobs arriving at the $i$-th station on their path {$\psi_r$} that are then lost at the $i$-th station. 
We express the performance of the above system in terms of the expected net rate of reward generation by the system operating in equilibrium. 
For both variants, increased losses result in reduced revenue for the network operator.  
The difference between the variants is in how rewards are collected by the network operator, which is described as follows. 
\begin{itemize}
\item For Model~I, once a customer of class $r$ has successfully entered all of the stations in its path $\psi_r$, the network operator receives revenue $\REV_r$. 
This means that in equilibrium with capacity $\boldsymbol c$ the network operator receives payments of value $\REV_r$ at rate $\lambda_r$ thinned by $\prod_{i=1}^{N_r} \big(1-p_{r,i}(\boldsymbol c)\big)$. 
The latter product represents the proportion of class-$r$ customers that are not lost at any of the stations in $\psi_r$. 
Summing these revenue streams across all of the classes {$r\in\mathcal R$ together} with the costs of providing servers $\langle \boldsymbol c, \boldsymbol \COST \rangle$ results in the objective function
\begin{equation}\label{eq:OBJ_1}
f(\boldsymbol c) = -\langle \boldsymbol c, \boldsymbol \COST \rangle +\sum_{r\in \mathcal R}\lambda_r\,\REV_{r}\prod_{i=1}^{N_r} \big(1-p_{r,i}(\boldsymbol c)\big)\,. 
\end{equation}
This expression serves as our Model~I objective function, which is inspired by the {\em capacity value function} investigated in \cite{Chiera2005,Chiera2002} within the context
of individual Erlang-B loss systems and later generalised in \cite{Patch2018transient} within the context of provisioning cloud computing platforms.
\item For Model~II, when a customer of class $r$ successfully enters station $l(r,i)$ along its path $\psi_r$, the network operator receives revenue $\REV_{r,i}$. 
In order for this revenue to be obtained, the customer must be lost at stations $l(r,1),\ldots, l(r,i-1)$ before being accepted at station $l(r,i)$. 
Hence, in equilibrium with capacity $\boldsymbol c$, the network operator receives payments of value $\REV_{r,i}$ at rate $\lambda_r$ thinned by $\big(1-p_{r,i}(\boldsymbol c)\big)\prod_{j=1}^{i-1}p_{r,j}(\boldsymbol c)$. 
This product represents the proportion of class-$r$ customers that are lost at stations $l(r,1),\ldots,l(r,i-1)$ and then accepted at station $l(r,i)$. 
Combining these revenue streams across all of the classes {$r\in\mathcal R$ together} with the costs of providing servers $\langle \boldsymbol c, \boldsymbol \COST \rangle$ results in the objective function
\begin{equation}\label{eq:OBJ_2}
f(\boldsymbol c) = -\langle \boldsymbol c, \boldsymbol \COST \rangle +\sum_{r\in \mathcal R} \lambda_r \sum_{i=1}^{N_r}\REV_{r,i}\big(1-p_{r,i}(\boldsymbol c)\big)\prod_{j=1}^{i-1}p_{r,j}(\boldsymbol c)\,. 
\end{equation}
This expression serves as our Model~II objective function. 
\end{itemize}
For both {model} variants, our optimisation formulation seeks to determine
\begin{equation}\label{eq:max}
\boldsymbol c^\star := \underset{\boldsymbol c\in\Ints_+^L}{\argmax}~f(\boldsymbol c)\,,
\end{equation}
{where $\Ints_+$ denotes the non-negative integers.}

\begin{figure}
 \begin{subfigure}[b]{.45\linewidth}
 \centering
  \includegraphics[width = 0.7\textwidth]{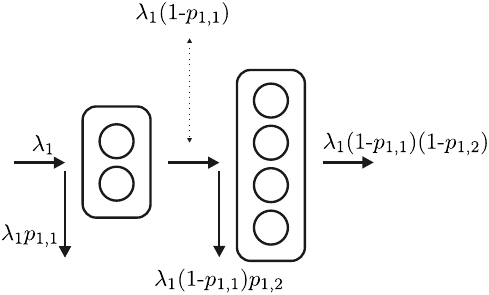}
  \caption{Tandem network Model~I.}
  \label{fig:network2_1}
  \end{subfigure}
 \begin{subfigure}[b]{.45\linewidth}
 \centering
  \includegraphics[width = 0.7\textwidth]{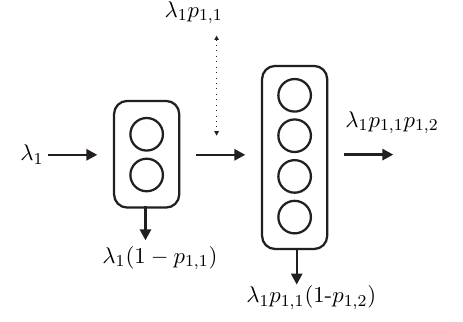}
  \caption{Tandem network Model~II.}
  \label{fig:network2_2}
  \end{subfigure}
  \caption{Two small network instances.}
  \label{fig:small_net}
\end{figure}

{\bf Example.}
In order to
{elucidate}
the two variants of {our general class of} networks of
SLSs,
we now briefly discuss a simple {illustrative} example.
Consider a system with two stations, a single customer class, and capacities $c_1$, $c_2$;. 
Figure~\ref{fig:network2_1} illustrates the Model~I variant of this system and Figure~\ref{fig:network2_2} illustrates the Model~II variant of this system. 
Customers arrive to station $1$ according to a Poisson process with rate $\lambda_1$;
upon successful service at station $1$ under Model~I, the customer attempts service at station 2, whereas upon a loss at station 1 under Model~II, the customer attempts service at station $2$.
Hence, $\mathcal L = \{1,2\}$, $\mathcal R = \{1\}$, $\psi_1=(1,2)$, $l(1,1)=1$, $l(1,2)=2$. 
Observe that, while $p_{1,1}(c_1)$ is independent of $c_2$, $p_{1,2}(\boldsymbol c)$ depends on both $c_1$ and $c_2$ because larger $c_1$ renders a larger {(smaller)} number of arrivals to station $2$
{under Model I (Model II)}.
Capacity costs are $\COST_1$ and $\COST_2$ per unit time.
In the case of Model~I, rewards $\REV_{1}$ are obtained by the network operator per successful entry at station $2$, hence the
{Model~I objective function}
\eqref{eq:OBJ_1} is given by
\begin{equation}\label{eq:TandemOBJ_1}
f(\boldsymbol c) = -\COST_1\,c_1-\COST_2\,c_2+\lambda_1\,\REV_{1}\,\big(1-p_{1,1}(c_1)\big)\big(1-p_{1,2}(\boldsymbol c)\big)\,. 
\end{equation}
In the case of Model~II, rewards $\REV_{1,1}$ and $\REV_{1,2}$ are obtained by the network operator per successful entry to stations $1$ and $2$ respectively, hence the
{Model~II objective function}
\eqref{eq:OBJ_2} is given by
\begin{equation}\label{eq:TandemOBJ_2}
f(\boldsymbol c) = -\COST_1\,c_1-\COST_2\,c_2+\lambda_1\,\REV_{1,1}\,\big(1-p_{1,1}(c_1)\big) +\lambda_1\,\REV_{1,2}\,\big(1-p_{1,2}(\boldsymbol c)\big)p_{1,1}(c_1)\,. 
\end{equation}
In both variants the capacity $\boldsymbol c$ is to be maximized over
a set of possible network parameters
$\mathcal C = \Ints_+^2$.

\section{Functional-form approach}\label{sec:FF4SNB}
We now
{present our general approximation approach that seeks to extend the first-phase}
approach in~\cite{Dieker2017} to our setting of networks of
SLSs.
A detailed description of our general approach and the underlying ideas is provided first,
and then we turn our attention to applying this approach in the specific setting of interest.

\subsection{Overview of general approach}\label{sec:opt}
The objective of our study is to determine parameters $\boldsymbol c$
that maximise a function
$f : \mathcal C \to \mathbb R$ representing a performance metric of interest.
In this paper we consider $\boldsymbol c$ to be the number of servers in a network of
SLSs
and $f(\boldsymbol c)$ to be a combination of the weighted ergodic fraction of lost customers
and the server provisioning costs.
We note that
Dieker et al.~\cite{Dieker2017} consider $f(\boldsymbol c)$ to be the weighted expected number of customers in a stochastic network (without losses) in equilibrium.

A primary difficulty in determining the value
\begin{equation}\label{eqn:opt}
\boldsymbol c^\star := \underset{\boldsymbol c\in\mathcal C}{\argmax}~f(\boldsymbol c)
\end{equation}
comes from the fact that the function $f$ is typically unknown, owing to the complexity of stochastic networks (especially in non-Markovian settings),
and therefore we assume it can only be evaluated as a random variable (r.v.) $F$ such that $\E F(\boldsymbol c) = f(\boldsymbol c)$.
With time-consuming simulation, we assume it is possible to obtain independent samples of $F$, denoted by $\hat f$, such that the information contained in these samples can be used to find an approximation of $\boldsymbol c^\star$.
The objective of this paper is to develop a method for closely approximating $\boldsymbol c^\star$ using as few evaluations of $F$ as possible,
for
{each of the model instances}
detailed in Section~\ref{sec:Model}.

Our {iterative} functional-form optimisation approach exploits underlying theoretical properties of $f$ (i.e., existing domain knowledge) to augment information from simulation and speed up
the iteration process of finding a good approximation to $\boldsymbol c^\star$.
This structural information is expressed as a closed-form function $\tilde{f}(\boldsymbol c,\boldsymbol\tau)$ of both the network parameters $\boldsymbol c$ and some additional (potentially vector-valued)
parameters $\boldsymbol\tau$.
The parameters $\boldsymbol\tau$ are then used to tune $\tilde f$ so that it fits $f$ well locally.
Depending on the complexity of $\tilde{f}$ we then solve, either analytically or numerically, for
\begin{equation}\label{eqn:approxOpt}
\tilde{\boldsymbol c}^\star = \underset{\boldsymbol c\in\mathcal C}{\argmax}~\tilde f(\boldsymbol c,\boldsymbol\tau)
\end{equation}
to approximate $\boldsymbol c^\star$.
The function $\tilde{f}$ is selected to ensure that~\eqref{eqn:approxOpt} can be solved in closed form or using a fast numerical procedure,
which is in stark contrast to solving for~\eqref{eqn:opt}.

The quality of the approximation \eqref{eqn:approxOpt}
relies in large part on choosing an $\tilde{f}$ that properly represents
fundamental behaviours of the stochastic network.
We assume $\boldsymbol\tau$ is selected from an appropriate set such that the set of functions $\big\{\tilde f(\cdot,\boldsymbol \tau)\big\}_{\boldsymbol \tau}$ has
elements that approximate $f$ well and that can be reliably identified in terms of $\boldsymbol\tau$.
Let us initially assume a good functional form is known {(which is addressed below in Section~\ref{sec:opt:appl})} and provide an iterative procedure for choosing $\boldsymbol\tau$,
where the sequence of samples
$\hat f(\boldsymbol c^{(1)}), \ldots, \hat f(\boldsymbol c^{(N)})$ 
is used to guide the selection of $\boldsymbol \tau$ over a maximum number of iterations $N$.

\begin{figure}
 \hspace{1cm}
  \begin{subfigure}[b]{.40\linewidth}
  \includegraphics[width = 0.70\textwidth]{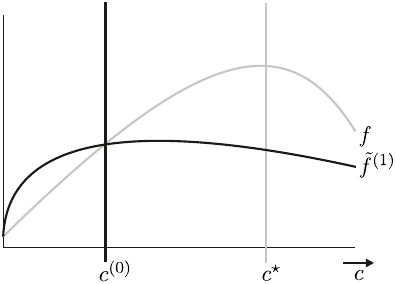}
  \caption{Step 1.}
  \label{fig:step1}
  \end{subfigure}
 \begin{subfigure}[b]{.40\linewidth}
  \includegraphics[width = 0.70\textwidth]{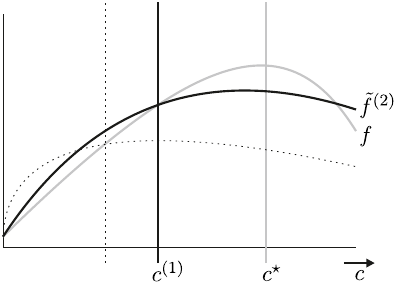}
  \caption{Step 2.}
  \label{fig:step2}
 \end{subfigure}
  \caption{Two iteration steps of a functional-form optimisation algorithm.}
 \label{fig:iteration_steps}
\end{figure}

Starting with an initial value of $\boldsymbol c=\boldsymbol c^{(0)}$,
which can be chosen at random {or via a crude analytical approximation}, 
we first evaluate $\hat f(\boldsymbol c^{(0)})$ using simulation.
Here $\hat f(\boldsymbol c^{(0)})$ represents the objective value at $\boldsymbol c=\boldsymbol c^{(0)}$ obtained through simulation.
The next step is to set $\hat f(\boldsymbol c^{(0)}) = \tilde f(\boldsymbol c^{(0)},\boldsymbol\tau^{(1)})$ and solve for $\boldsymbol\tau^{(1)}$.
This renders our iteration~$1$ approximation function $\tilde f^{(1)}(\cdot) := \tilde f(\cdot,\boldsymbol\tau^{(1)})$,
as depicted in Figure~\ref{fig:step1}\footnote{Although {Figure~\ref{fig:iteration_steps}} depicts the domain of $f$ to be one-dimensional for ease of illustration,
we are generally interested in high-dimensional stochastic networks, which represent a primary cause of complexity for the problems under consideration.}.
Then, from \eqref{eqn:approxOpt}, we solve $\boldsymbol c^{(1)} := \argmax_{\boldsymbol c}\,\tilde f^{(1)}(\boldsymbol c)$ to obtain our iteration~$1$
approximation for $\boldsymbol c^\star$.

Continuing in this manner, we evaluate $\hat f(\boldsymbol c^{(1)})$ via simulation and solve
for $\boldsymbol\tau^{(2)}$ in $\hat f(\boldsymbol c^{(1)}) = \tilde{f}(\boldsymbol c^{(1)},\boldsymbol\tau^{(2)})$
to obtain our iteration~$2$ approximation $\tilde{f}^{(2)}(\cdot) := \tilde{f}(\cdot,\boldsymbol\tau^{(2)})$ that
intersects $f$ at $\boldsymbol c = \boldsymbol c^{(1)}$ in expectation; see Figure~\ref{fig:step2}.
Observe that, while selection of the functional form $\tilde{f}$ requires fundamental insights into the stochastic network of interest,
our approach does not rely solely on inaccurate queueing formulas, in contrast to purely analytic approximations, because simulation is used to evaluate the stochastic network.

In general, our iteration process consists of the sequence of approximation functions
\begin{equation}\label{eqn:alg1}
\tilde{f}^{(n)}(\cdot) := \tilde{f}(\cdot,\boldsymbol\tau^{(n)}), \quad n=1,2,\dots\,,
\end{equation}
with corresponding maximisers
\begin{equation}
\boldsymbol c^{(n)} := \underset{\boldsymbol c\in\mathcal C}{\argmax}~\tilde{f}^{(n)}(\boldsymbol c), \quad n=1,2,\dots\,,
\end{equation}
where the tuning parameter $\boldsymbol\tau^{(n)}$ in each iteration is obtained by solving
\begin{equation}\label{eqn:alg3}
\hat f(\boldsymbol c^{(n-1)}) := \tilde{f}(\boldsymbol c^{(n-1)},\boldsymbol\tau^{(n)}), \quad n=1,2,\dots\,.
\end{equation}
We then iterate \eqref{eqn:alg1}~--~\eqref{eqn:alg3} until the difference $||\boldsymbol c^{(n)}-\boldsymbol c^{(n-1)}||$
is sufficiently small or a maximum number of iterations $N$ has been reached. 

{Our general}
approach relies on the fact that $\tilde{f}(\cdot,\boldsymbol\tau^{(n)})$ provides a good approximation for $f$ around $\boldsymbol c=\boldsymbol c^{(n-1)}$ and
thus the algorithm is likely to move in
{a}
proper direction at each iteration.
This also provides significant computational improvements over purely simulation-based methods:
Instead of, for instance, running many expensive simulations to estimate the Jacobian (or Hessian) as in
{SA},
our {iterative} functional-form approach
essentially uses $\tilde{f}$ (a function $\mathcal C \to \mathbb R$) to provide $\nabla \tilde f(\boldsymbol c)$ as an approximation for the gradients
$\nabla f(\boldsymbol c)$, which requires only a single evaluation of $\hat f$ per iteration.

Following {the two-phase approach in}~\cite{Dieker2017}, when
{the final solution $\boldsymbol c^{(n)}$ from our our general first-phase approximation approach}
needs to be further refined,
{then}
$\boldsymbol c^{(n)}$
can be used as a {high-quality} starting point for a
simulation-based optimisation approach with guaranteed convergence properties, such as
{SA}.
{Such a second-phase solution}
may
{provide}
improved accuracy over our {general first-phase} approach above, but it also
{incurs additional computational costs}.
However, by first obtaining a near-optimal solution using our fast {iterative} functional-form approach, the more expensive purely simulation-based approach requires
far fewer iterations to find the optimal solution, thus significantly reducing the overall computational costs over exclusively using simulation-based optimisation~\cite{Dieker2017}.
{The solution of our first-phase approximation approach therefore can be used directly at finer time scales or indirectly at coarser time scales as a high-quality initial starting point for a preferred simulation-based algorithm.}

\subsection{Application to networks of stochastic loss systems}
\label{sec:opt:appl}
We aim to solve the resource optimisation problem~\eqref{eq:max} by determining the server capacity $\boldsymbol c\in\Ints_+^L$ that maximises objective function \eqref{eq:OBJ_1} or \eqref{eq:OBJ_2}. 
In most cases of interest, we are unable to do so directly because these objective functions must be evaluated via simulation.  
The costs $\boldsymbol \COST$, rewards $\boldsymbol\REV$ and external arrival rates $(\lambda_r)_{r\in\mathcal R}$ are assumed to be known constant vectors,
with each $\lambda_r$ independent of the resource allocation $\boldsymbol c$.
Recall that $p_{r,i}(\boldsymbol c)$ denotes the proportion of class-$r$ customers arriving at station $l(r,i)$ (i.e., $i$-th station of their path $\psi_r$) that are then lost at $l(r,i)$
as a function of $\boldsymbol c$.
The function $p_{r,i}(\boldsymbol c)$ depends on the long-term behaviour of the underlying stochastic process and is not explicitly known in general. 

The complexity of $f$ in \eqref{eq:OBJ_1} or \eqref{eq:OBJ_2} is contained in the set of functions $\{ p_{r,i} \,: \,r \in\mathcal R,\,i\in\mathcal L\}$, for which no closed-form expression is known. 
We therefore propose to replace each function $p_{r,i}$ with an approximation $\tilde p(\cdot, \tau_{r,i})$ that follows a functional form which incorporates features
that can be
reasonably believed to be present in $p_{r,i}$.
Here $\tau_{r,i}$ is an unknown parameter of the functional form indexed by class $r$ and station $i$.
Define $p_{r,0}(\boldsymbol c) \equiv 0$.
For networks of
SLSs
with objective function \eqref{eq:OBJ_1} and
$p_{r,i-1}(\boldsymbol c)<1$,
it is reasonable to expect that:
(i) $p_{r,i}(\boldsymbol c)$ is strictly decreasing in $c_{l(r,i)}$;
(ii)
$p_{r,i}(\boldsymbol c) = 1$ when $c_{l(r,i)}=0$; and
(iii) $p_{r,i}(\boldsymbol c) \to 0$ as $c_{l(r,i)} \to \infty$. 
Expectations (i)--(iii) similarly hold for networks with objective function \eqref{eq:OBJ_2} and
$p_{r,i-1}(\boldsymbol c)>0$.
By definition,
the
case of
$p_{r,i-1}(\boldsymbol c)=1$ 
for objective \eqref{eq:OBJ_1}
implies $p_{r,i}(\boldsymbol c)=0$ for all $c_{l(r,i)}$
(given no class $r$ arrivals to station $l(r,i)$);
similarly,
by definition, the case of
$p_{r,i-1}(\boldsymbol c)=0$ for objective \eqref{eq:OBJ_2} implies $p_{r,i}(\boldsymbol c)=0$ for all $c_{l(r,i)}$. 
In these instances,
changes to $c_{l(r,i)}$ do not affect the objective function via $p_{r,i}(\boldsymbol c)$. 
We therefore focus on cases where $c_{l(r,i)}$ influences the objective function by finding an approximation $\tilde p(\cdot,\tau_{r,i})$ for $p_{r,i}$ that satisfies properties (i)--(iii). 
Since
these properties
depend only on the resource allocation $c_{l(r,i)}$ at the relevant station $l(r,i)$,
we take $\tilde p(\cdot,\tau_{r,i})$ to be a univariate function of $c_{l(r,i)}$ (disregarding other components of $\boldsymbol c$). 
We observe that cumulative probability distribution functions for r.v.s with non-negative support form a class of functions with exactly these properties. 
Many such candidate functions were tested,
including but not limited to (in the one-dimensional case) $\exp(-(c\tau)^k)$ and $1/(1+(c\tau)^k)$ for a wide range of $k$, as well as $(\tau^{c}/c!)\left(\sum_{i=0}^{c} \tau^i/i!\right)^{-1}$.
Due to space constraints, we
do not present results on all of these forms but instead
focus on
the functional form
found to work best for {the} networks of
SLSs
{of interest}:
\begin{equation}\label{eqn:weibull}
\tilde p(c,\,\tau_{r,i}) = \exp\left(-(c\,\tau_{r,i})^2\right)\,.
\end{equation}

As we will see in Section~\ref{sec:Experiments}, the above choice indeed performs very well.
{Although the}
capacity values in \eqref{eq:max} are discrete,
{the}
application of our optimisation approach (described below) using \eqref{eqn:weibull} requires continuous functions of $\boldsymbol c$.
We therefore apply the following assumption during the optimisation procedure. 
\begin{assumption}\label{assume1}
When a customer arrives to a station with (non-integer) capacity $c$ and finds $\lfloor c \rfloor+1$ customers, the customer is lost.
If an arrival encounters $\lfloor c \rfloor$ customers, the newly arriving customer is accepted with probability $c-\lfloor c \rfloor$, and lost otherwise.
When an arrival finds less than or equal to $\lfloor c \rfloor-1$ customers, the arrival is accepted.
\end{assumption}
This assumption results in a continuous relaxation of the objective functions that matches \eqref{eq:OBJ_1} or \eqref{eq:OBJ_2} at integer points.
Whenever our (approximate) optimal solution is non-integer, it is rounded to the closest integer.
We note that it is not uncommon to use a continuous relaxation of a discrete problem in this way (see, e.g., \cite[Section~3.1]{amaran2016simulation}). 

Our algorithm is formulated following \eqref{eqn:alg1}~--~\eqref{eqn:alg3} and formally presented in Algorithm~\ref{ALG1final} below. 
Let $\hat p_{r,i}(\boldsymbol c)$ denote an unbiased estimator of $p_{r,i}(\boldsymbol c)$ obtained from simulation for an allocation $\boldsymbol c$.
Given an initial guess (possibly at random) for the optimal capacity allocation $\boldsymbol c^{(0)}$, we simulate the system to obtain the loss probabilities $\hat p_{r,i}(\boldsymbol c^{(0)})$
for each $(r,i)$ pair of class and station indices. 
Using this we can determine the parameters $\tau_{r,i}^{(1)}$ by solving the implicit equations
\begin{equation}\label{eqn:tau_gen}
\hat p_{r,i}(\boldsymbol c^{(0)}) = \tilde p\left(c_{l(r,i)}^{(0)},\,\tau_{r,i}^{(1)}\right)\,.
\end{equation}
Namely, we want to calibrate our approximation $\tilde p(\cdot, \tau_{r,i}^{(1)})$ by choosing $\tau_{r,i}^{(1)}$ such that the predicted loss probability at $\boldsymbol c^{(0)}$ matches the observed loss probability. 
After some {algebraic} manipulation we have
$$
\tau_{r,i}^{(1)} = \frac{1}{c_{l(r,i)}^{(0)}}\Big(-\log\big(\hat p_{r,i}\left(\boldsymbol c^{(0)}\right)\big)\Big)^{1/2}.
$$

Once these $\tau_{r,i}^{(1)}$ are determined, for Model~I we can obtain a new approximation to the optimal resource allocation problem by evaluating the corresponding version
of \eqref{eq:OBJ_1} and \eqref{eq:max}:
\begin{equation}\label{eq:OPT1approx}
\underset{\boldsymbol c^{(1)}\in \mathbb [0,\infty)^L}{\argmax} \quad-\langle \boldsymbol c^{(1)}, \boldsymbol \COST \rangle+\sum_{r\in \mathcal R}\lambda_r\,\REV_{r} \,\prod_{j=1}^{i} \left(1-\tilde p\big(c_{l(r,i)}^{(1 )},\,\tau_{r,i}^{(1)}\big) \right)\,,
\end{equation}
with $\tilde p(\cdot, \tau_{r,i})$ given by \eqref{eqn:weibull}. 
Solving~\eqref{eq:OPT1approx}, which can be done efficiently using standard optimisation techniques, renders a new resource allocation, which is used as the basis for the next step. 
Similarly, for Model~II, once the collection of $\tau_{r,i}^{(1)}$ are determined we can obtain a new approximation to the optimal resource allocation problem by evaluating the corresponding version
of \eqref{eq:OBJ_2} and \eqref{eq:max}:
\begin{equation}\label{eq:OPT2approx}
\underset{\boldsymbol c^{(1)}\in \mathbb [0,\infty)^L}{\argmax} \quad-\langle \boldsymbol c^{(1)}, \boldsymbol \COST \rangle +\sum_{r\in \mathcal R} \lambda_r \sum_{i=1}^{N_r}\REV_{r,i}\Big(1-\tilde p\big(c_{l(r,i)}^{(1 )},\,\tau_{r,i}^{(1)}\big)\Big)\prod_{j=1}^{i-1}\tilde p\big(c_{l(r,j)}^{(1 )},\,\tau_{r,j}^{(1)}\big)\,,
\end{equation}
with $\tilde p(\cdot, \tau_{r,i})$ given by \eqref{eqn:weibull}.
Solving~\eqref{eq:OPT2approx} yields a new resource allocation, which is used as the basis for the next step. 
Iteratively applying these steps under either Model~I or Model~II for a desired level of accuracy $\epsilon>0$ or for a maximum number of steps $N\in \Ints^+$ leads to Algorithm~\ref{ALG1final}. 
This algorithm has (at least) two advantages over other optimisation techniques: (i) it explicitly incorporates the functional form of the objective function \eqref{eq:OBJ_1} or \eqref{eq:OBJ_2}; and (ii) the functional-form approximation $\tilde p(c,\,\tau_{r,i}) = \exp\left(-(c\,\tau_{r,i})^2\right)$ given in \eqref{eqn:weibull} explicitly utilises known fundamental features of the intractable quantities $p_{r,i}(\boldsymbol c)$.

\begin{algorithm}
\SetAlgoLined
\KwResult{Approximation to optimal capacity allocation.}
Choose $\boldsymbol c^{(0)}\in [1,\infty)^L,\,\epsilon>0,\,N\in\Ints^+$\;
Set $n=1$ and $g > \epsilon$\;
\While{$g >\epsilon$ and $n\le N$}{
 	For $r\in\mathcal R$ and
{$i\in [N_r]$,}
evaluate $\hat p_{r,i}(\boldsymbol c^{(n-1)})$  using simulation\;
 	For $r\in\mathcal R$ and
{$i\in [N_r]$,}
set $\tau_{r,i}^{(n)} = \frac{1}{c_{l(r,i)}^{(n-1)}}\left(-\log\left(\hat p_{r,i}\left(c_{l(r,i)}^{(n-1)}\right)\right)\right)^{1/2}$\;
	To approximate the optimal capacity for Model~I, numerically solve
\begin{equation}\label{eq:ALG1}
\boldsymbol c^{(n)} = \underset{\boldsymbol c\in \mathbb [0,\infty)^L}{\argmax} \quad-\langle \boldsymbol c, \boldsymbol \COST \rangle+\sum_{r\in \mathcal R}\lambda_r\,\REV_{r} \,\prod_{i=1}^{N_r} \Big(1-\exp\left(-\big(c_{l(r,i)}\tau_{r,i}^{(n)}\big)^2\right)\Big)\,,
\end{equation}
or for Model~II, numerically solve
\begin{equation}\label{eq:ALG2}
\boldsymbol c^{(n)} = \underset{\boldsymbol c\in \mathbb [0,\infty)^L}{\argmax} \quad-\langle \boldsymbol c, \boldsymbol \COST \rangle +\sum_{r\in \mathcal R} \lambda_r \sum_{i=1}^{N_r}\REV_{r,i}\Big(1-\exp\left(-\big(c_{l(r,j)}\tau_{r,i}^{(n)}\big)^2\right)\Big)\prod_{j=1}^{i-1}\exp\left(-\big(c_{l(r,j)}\tau_{r,j}^{(n)}\big)^2\right)\,. 
\end{equation}
Set $g = ||\boldsymbol c^{(n-1)}-\boldsymbol c^{(n)}||$ and $n=n+1$\;
 }
  Output $\boldsymbol c^{(n)}$.
 \caption{Functional-form optimisation. }
 \label{ALG1final}
\end{algorithm}

Assuming a maximum number of servers available for allocation, \eqref{eq:OPT1approx} and \eqref{eq:OPT2approx} define a mapping for which the existence of a fixed point can be established using Brouwer's theorem. 
Showing that this is also a contraction mapping (and therefore possesses a unique fixed point) can be very difficult because doing so relies on using properties of $p_{r,i}(\boldsymbol c)$ which are not known in general. 
Even in the simple case of a single station with one class, Poisson arrivals and exponential service times,
showing that the mapping is a contraction involves handling highly complicated expressions in terms of the derivatives (with respect to $c$) of the function 
\[
\frac{\lambda^{c}/c!}{\sum_{i=0}^{c} \lambda^i/i!}\,. 
\]
{For these reasons, we leave for future work the problem of showing that Algorithm~\ref{ALG1final} provably converges},
{noting that this currently remains}
an (highly challenging) open problem. 
By the {inherent} nature of the problem at hand, the efficiency of any implementation of Algorithm~\ref{ALG1final} will be dominated by the speed with which the simulation samples $\hat p(\boldsymbol c)$ can be generated. 
Our algorithm takes as input a vector of size $L$ and then in each iteration, after generating the set of estimates $\hat p_{r,i}(\boldsymbol c)$, the quantities $\tau_{r,i}$ must be computed for each class-station pair. 
Relative to obtaining the estimates $\hat p_{r,i}$, computation of $\tau_{r,i}$ is very efficient, as is solving the optimisation problem in Step~6.


\section{Computational experiments}\label{sec:Experiments}
\begin{figure}
\centering
\hspace{1.5cm}
 \begin{subfigure}[b]{.4\linewidth}
  \hspace{10mm}\includegraphics[width = 0.7\textwidth]{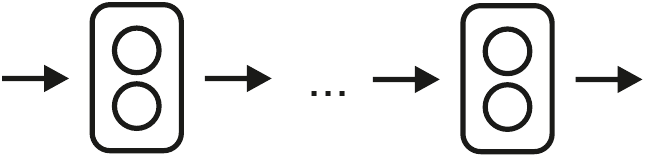}
  \vspace{5mm}
  \caption{Tandem network.}
  \label{fig:top1}
 \end{subfigure}
 \begin{subfigure}[b]{.4\linewidth}
  \hspace{10mm}\includegraphics[width = 0.7\textwidth]{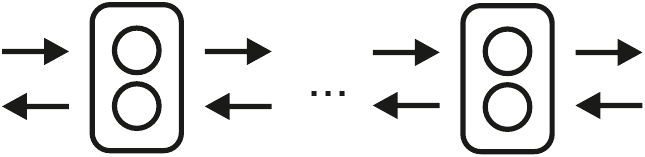}
    \vspace{5mm}
    \caption{Criss-cross network. }
  \label{fig:top2}
 \end{subfigure}
\vspace{5mm}

  \begin{subfigure}[b]{.4\linewidth}
  \hspace{10mm}\includegraphics[width = 0.75\textwidth]{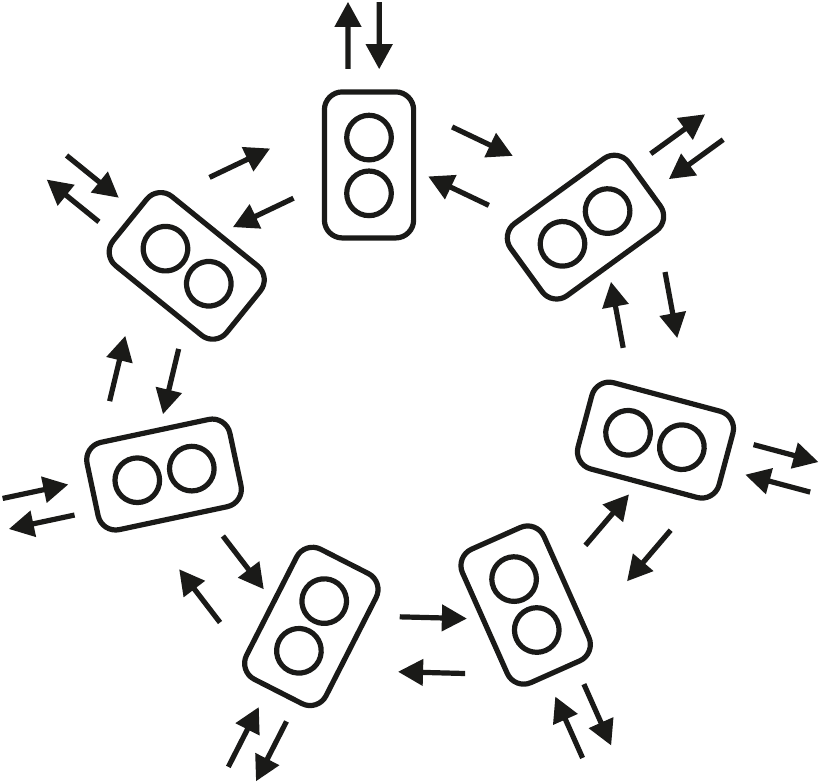}
    \caption{Ring network.}
  \label{fig:top3}
 \end{subfigure}
  \begin{subfigure}[b]{.4\linewidth}
  \includegraphics[width = 0.75\textwidth]{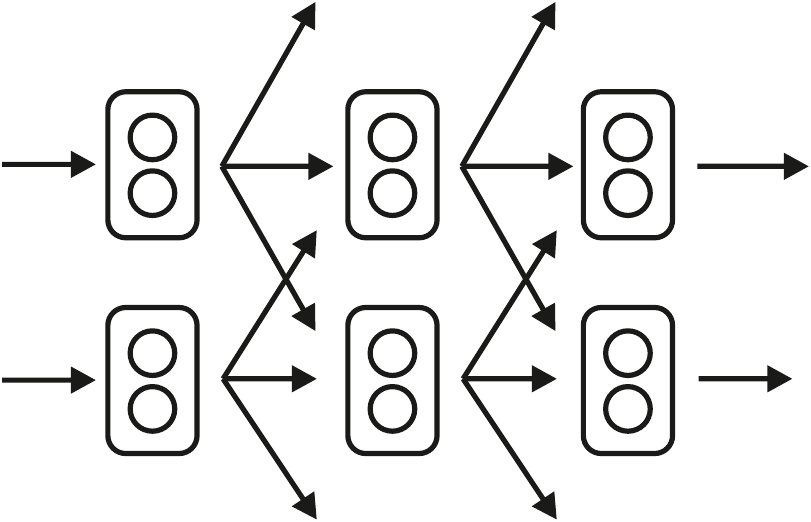}
  \vspace{5mm}
    \caption{Canonical network.}
  \label{fig:top4}
 \end{subfigure}
  \caption{Topologies of networks for computational experiments.}
\label{fig:tops}
\end{figure}

In this section 
{we perform extensive computational experiments to investigate and compare the performance of the core iterative algorithm of our first-phase approximation approach, reporting and discussing}
experimental results for four example
networks of
SLSs:
(a) a Markovian tandem network; (b) a tandem network with cross-routes; (c) a ring network; and (d) a canonical network. 
These network topologies are
{illustrated}
in Figure~\ref{fig:tops}. 
The goal of the experiments is to gauge and compare the accuracy and efficiency of our {core} algorithm, presented in the previous section, against benchmark algorithms and results.
To our knowledge there are no examples of other algorithms tailored specifically to optimise the
{networks of
SLSs
of interest herein}.
For the Markovian tandem network example, we are able to derive explicit matrix-analytic expressions for the objective function (see Appendix~\ref{app:MAM}) that provide a benchmark for accuracy. 
Such expressions are not
{available}
for the other three example topologies of general networks of
SLSs.
For this reason, we compare the accuracy and efficiency of our {core iterative} algorithm against three benchmark algorithms: (i) a fine-tuned {SA} implementation (see Appendix~\ref{app:StochApprox});
(ii) {SO} (as implemented in Matlab 2020b with the {\sf surrogateopt} function); and (iii) {BO} (as implemented in Matlab 2020b with the {\sf bayesopt} function). 

In each of the experiments, all the algorithms are based on mean-value Monte Carlo simulations of the models. 
Given a capacity allocation $\boldsymbol c$, the system is simulated from an initial empty state to obtain $\hat p_{r,i}(\boldsymbol c)$. 
Time is then batched into prespecified periods of length $\tilde T$.
At the end of each time period, approximate $95$\% confidence interval
{widths}
for all $\hat p_{r,i}(\boldsymbol c)$ are computed using the score method for Bernoulli-type data
(see e.g., \cite[Example~6.16]{kroese2014statistical}) as follows:
\begin{equation}\label{eq:CIgap}
\frac{1.96 \sqrt{1.96^2-4 A^{(n)}_{r,i} (\hat p_{r,i}(\boldsymbol c)-1)\hat p_{r,i}(\boldsymbol c)}}{1.96^2+A^{(n)}_{r,i}}\,,
\end{equation}
where $A^{(n)}_{r,i}$ is the number of arrivals to station-class pair $(r, i)$ in time period $n$. 
These approximate confidence intervals are appropriate since each arrival to a station can be viewed as a Bernoulli trial with parameter representing the probability that the arrival is accepted for processing. 
When all of these confidence interval
{widths}
fall below a specified level of $0.01$ or a maximum simulation clock is reached, the simulation ends and a value is returned.  
In our experiments, we follow this procedure $10$ times and then return the average of these values as our estimates $\hat p_{r,i}(\boldsymbol c)$.  
After extensive tuning, we found that setting the maximum simulation clock to $100$ provided a good balance between speed and accuracy (so that our numerous experiments could be completed within a reasonable time-frame). 
Some capacity allocations result in the true loss proportion being extremely close to $0$, which can result in an inaccurate estimate of $0$ being returned with high probability. 
Since our model does not allow for loss proportions with value $0$ unless capacity is also $0$,
we set the estimate $\hat p_{r,i}(\boldsymbol c)$ to $10^{-6}$ whenever a loss proportion of $0$ is estimated with positive capacity.
There are a variety of other ways that these estimates could be obtained, which could also be used within our optimisation framework, but this method was chosen for its simplicity. 

The computational experiments are performed in Matlab 2020b and the numerical optimisation package to implement Step~6 of Algorithm~\ref{ALG1final} is the interior point algorithm built into this platform. 
The experiments were conducted on an Intel Core i7-3770 3.4GHz workstation running Windows 10 with 16GB of RAM.  
Code for these experiments is available online at \href{https://github.com/bpatch/networks-of-stochastic-loss-systems}{https://github.com/bpatch/networks-of-stochastic-loss-systems}, where we also provide data files containing the results of completed experiments. 
All experiments are conducted with a fixed seed (included in the code) so that they can be easily replicated. 
Due to the large number of parameters needed to generate all of the scenarios we consider, it is impractical to provide all of these parameters in table form here and therefore the interested reader is referred to the data files provided online. 
{In all of the box plots in this paper, outliers are defined as points that are greater than $q_{75}+w(q_{75}-q_{25})$ or less than $q_{25}-w(q_{75}-q_{25})$,
where $q_{n}$ is the $n$-th percentile and $w$ is approximately $2.7$ times the sample standard deviation (which corresponds to 99.3\% coverage, if the data are normally distributed).}

\subsection{Markovian tandem networks}\label{sec:TandemExperiments}
{\em Tandem Experiment 1 (illustrative).} 
We start with an experiment that illustrates how our algorithm and benchmarks evolve as they search for the optimal solution in a specific problem instance. 
Consider the example from Section~\ref{sec:Model}, displayed in Figure~\ref{fig:small_net}, with arrival rate $\lambda_1 = 16$, capacity costs $\COST_1=0.2$ and $\COST_2 = 0.3$,
and suppose that the service requirements of jobs are exponentially distributed with rate $0.8$ at the first station and $0.6$ at the second station. 
We consider both variants of our model: 
For Model~I, let rewards for successful service completions be $\REV_{1}=1.9$; and for Model~II, let $\REV_{1,1}=1$ and $\REV_{1,2}=0.9$. 

\begin{figure}[h]
  \begin{subfigure}[b]{0.2\textwidth}
    \centering
 \includegraphics{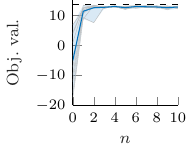}
  \caption{Alg.~1, Model~I.}
  \end{subfigure}
 \begin{subfigure}[b]{0.2\textwidth}
   \centering
 \includegraphics{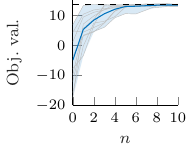}
  \caption{SA, Model~I.}
 \end{subfigure}
 \begin{subfigure}[b]{0.2\textwidth}
   \centering
 \includegraphics{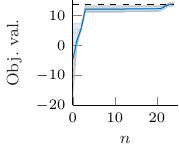}
  \caption{SO, Model~I.}
  \end{subfigure}
 \begin{subfigure}[b]{0.2\textwidth}
   \centering
 \includegraphics{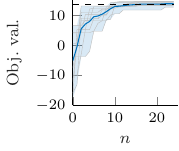}
  \caption{BO, Model~I.}
  \end{subfigure}

  \begin{subfigure}[b]{0.2\textwidth}
    \centering
 \includegraphics{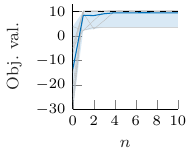}
  \caption{Alg.~1, Model~II.}
  \end{subfigure}
   \begin{subfigure}[b]{0.2\textwidth}
     \centering
 \includegraphics{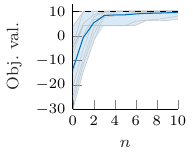}
  \caption{SA, Model~II.}
 \end{subfigure}
 \begin{subfigure}[b]{0.2\textwidth}
   \centering
 \includegraphics{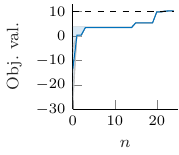}
  \caption{SO, Model~II.}
  \end{subfigure}
   \begin{subfigure}[b]{0.2\textwidth}
   \centering
 \includegraphics{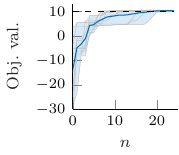}
  \caption{BO, Model~II.}
  \end{subfigure}
 \caption{Quality of solution obtained by Algorithm~\ref{ALG1final} and benchmarks for a tandem network (Tandem Experiment 1).}
 \label{fig:TendemAcc}
\end{figure}

For this simple network, it is possible to explicitly evaluate the objection functions \eqref{eq:TandemOBJ_1} and \eqref{eq:TandemOBJ_2} using matrix-analytic methods (MAMs) and then search for the optimal solution
{in a brute-force manner.}
Although this approach does not extend to more complex settings, we present these derivations in Appendix~\ref{app:MAM} to enable an exact evaluation of the accuracy of {Algorithm~\ref{ALG1final} compared to the SA, SO and BO benchmarks}. 
Computations from the MAM approach for Model~I provide the true maximum $f(\boldsymbol c^\star) = 13.4975$ at $\boldsymbol c^\star = (26,32)$.
On the other hand, for Model~II, the true maximum is $f(\boldsymbol c^\star) =10.2049$ at $\boldsymbol c^\star = (26,0)$. 
Figure~\ref{fig:TendemAcc} displays trajectories of the values returned by Algorithm~\ref{ALG1final}, SA, SO and BO as these algorithms search for the optimal function values $13.4975$ and  $10.2049$,
where the true maximum is displayed as a dashed line and the mean of the trajectories is displayed as a solid blue line,
together with a light blue shading that encompasses all trajectories; and each of the trajectories displayed in grey.
The displayed trajectories are from noisy evaluations of the objective function (and thus it is possible that they exceed the true maximum). 
Each trajectory starts from a random initial condition (which is held constant across approaches). 
{Our key observations for the results in this figure, which will be remarked upon soon, are as follows.}
\begin{itemize}
\item[-] Within the allocated number of iterations, all four methods are able to greatly improve upon the objective function value from that of the initial random capacity allocation. 
Specifically, sample paths of Algorithm~1 for both models achieve on average 95\% of the potential improvement that is possible from the given initial conditions;
one sample path under Model~II exhibits an improvement of only 56\%, suggesting this this trajectory has become stuck in a local solution. 
The objective function values returned by SO and BO consistently achieve 100\% of the potential improvement. 
The objective function values returned by SA were also highly competitive, respectively achieving 96.5\% and 99.7\% of the potential improvements for Model~1 and Model~II. 
\item[-] Trajectories of Algorithm~1 typically make an immediate large jump towards the optimal function value
(for Model~I, eight of the trajectories make a gain of more than $88\%$ by $n{=}2$; and for Model~II, eight of the trajectories make a gain of more than $99\%$ by $n{=}2$),
{much more quickly than the other methods},
and then they experience minor oscillations (less than $1$ and $0.5$ after $n{=}3$ for Model~I and Model~II, respectively).
\item[-] Trajectories of SA and BO make consistent but small movements towards the optimal function value,
whereas SO experiences mid-sized jumps towards the optimal function value separated by long periods of no movement (sometimes for more than $10$ iterations).
\end{itemize}

\begin{figure}[h]
  \begin{subfigure}[b]{0.2\linewidth}
    \centering
 \includegraphics{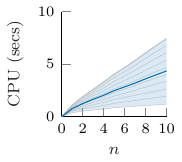}
  \caption{Alg.~1, Model~I.}
  \end{subfigure}
 \begin{subfigure}[b]{0.2\linewidth}
   \centering
 \includegraphics{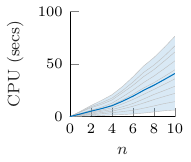}
  \caption{SA, Model~I.}
 \end{subfigure}
 \begin{subfigure}[b]{0.2\linewidth}
   \centering
 \includegraphics{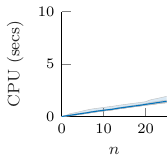}
  \caption{SO, Model~I.}
  \end{subfigure}
 \begin{subfigure}[b]{0.2\linewidth}
   \centering
 \includegraphics{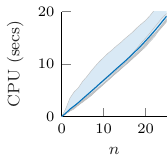}
  \caption{BO, Model~I.}
  \end{subfigure}

  \begin{subfigure}[b]{0.2\linewidth}
    \centering
 \includegraphics{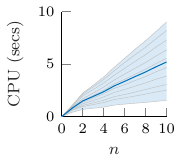}
  \caption{Alg.~1, Model~II.}
  \end{subfigure}
 \begin{subfigure}[b]{0.2\linewidth}
   \centering
 \includegraphics{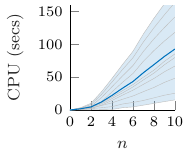}
  \caption{SA, Model~II.}
 \end{subfigure}
 \begin{subfigure}[b]{0.2\linewidth}
   \centering
 \includegraphics{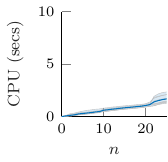}
  \caption{SO, Model~II.}
  \end{subfigure}
   \begin{subfigure}[b]{0.2\linewidth}
   \centering
 \includegraphics{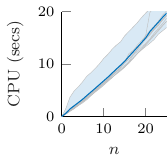}
  \caption{BO, Model~II.}
  \end{subfigure}
  \caption{Efficiency of Algorithm~\ref{ALG1final} and benchmarks for a tandem network (Tandem Experiment~1).}
 \label{fig:TendemEff}
\end{figure}

We follow up on our observations of the solution quality from the four algorithms with an investigation into the computing power needed to obtain these solutions. 
Figure~\ref{fig:TendemEff} displays the cumulative CPU time required to generate the objective values displayed in Figure~\ref{fig:TendemAcc} for each trajectory (grey) and on average (blue).  
{Our key observations for the results in this figure are as follows.}
\begin{itemize}
\item[-] {On average, to reach an objective function value within $1\%$ of the optimal value returned,
the CPU times for SA (Model~I: 3.6060 seconds; Model~II: 6.8583 seconds) and for BO (Model~I: 10.2423 seconds; Model~II: 9.0879 seconds) are substantially greater than
the CPU times for Algorithm~1 (Model~I: 0.2321 seconds; Model~II: 0.1493 seconds) and for SO (Model~I: 1.29 seconds; Model~II: 1.132 seconds),
with those for SO significantly greater than those for Algorithm~1.
Specifically, in comparison with Algorithm~1 on Model~I and Model~II,
SA requires a multplicative factor of $15.5$ and $45.9$ more CPU time, 
BO requires a multplicative factor of $44.1$ and $60.9$ more CPU time, and
SO requires a multplicative factor of $5.6$ and $7.6$ more CPU time.}
\item[-] The best solution for Model~I returned by Algorithm~\ref{ALG1final}, SA, SO and BO in these experiments were $(23,30)$, $(25,30)$, $(24,30)$ and $(26,30)$, respectively. 
For Model~II the best solution for Algorithm~1, SA and BO was $(26,0)$, while the best solution returned by SO was $(27,0)$. 
\end{itemize}

One important takeaway from this experiment is that all four algorithms are capable of reliably generating solutions which are close or equal to the optimal solution with very small optimality gaps. 
The optimal function values achieved by SO and BO were the best, but were followed closely by Algorithm~1 and SA.
Our algorithm {was} able to do so in the most efficient manner, and at speeds which were more than $5$-$7$ times faster than the next competitive method, namely SO.
This experiment also suggests that SA and BO are vastly less efficient.
For Model~II, we observed that our algorithm can become trapped in a local solution, something that did not occur for the benchmarks in this experiment. 
At this point, based on the initial simple experiment, it appears that our algorithm may offer vast speed improvements but may suffer slightly in terms of accuracy,
suggesting that SO may be best when accuracy is most important and Algorithm~1 is best when efficiency is most important. 
On the other hand, as previously discussed, the solution from our algorithm can be used as a high-quality starting point for SO
and then benefit from both accuracy and efficiency. 

{\em Tandem Experiment 2 (scenarios).} 
To investigate the accuracy of our algorithms more generally,
we now consider a broader range of randomly chosen parameters for each model variant of the Markovian tandem network displayed in Figure~\ref{fig:small_net}.
For Model~I, we study $20$ different randomly sampled scenarios each corresponding to different parameter values, where
the rewards, service rates, arrival rate and capacity costs are all sampled uniformly at random from $(0,5)$, $(0,3)$, $(0,20)$ and $(0.1,0.3)$, respectively.
Similarly, for Model~II, we sampled $20$ additional scenarios according to the same scheme, except with the rewards and capacity costs sampled uniformly at random from $(0,3)$ and $(0,0.4)$, respectively.
To ensure the MAM-based computations for finding the exact solution (Appendix~\ref{app:MAM}) did not become too burdensome,
we only experimented with scenarios where the optimal value as determined by these exact methods is an element of $[1, 60)^2$.
We conducted $10$ computational experiments for each scenario where each experiment was performed with a different randomly selected initial condition (which is held constant across methods). 

\begin{figure}[h!]
\centering
    	\setcounter{subfigure}{0}
\begin{subfigure}[b]{0.99\linewidth}
 \begin{subfigure}[b]{.33\linewidth}
 \centering
  \includegraphics{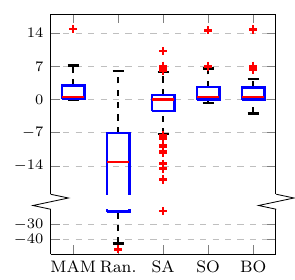}
  \caption{Difference between function value returned and value returned by Alg.~1. \\}
  \label{fig:tandem_obj-a}
  \end{subfigure}
  \hspace{2mm}
 \begin{subfigure}[b]{.33\linewidth}
 \centering
  \includegraphics{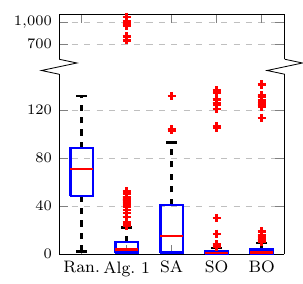}
  \caption{Euclidean distance between solution returned and true optimal solution.\\}
  \label{fig:tandem_sol-a}
  \end{subfigure}
    \hspace{2mm}
   \begin{subfigure}[b]{.33\linewidth}
 \centering
  \includegraphics{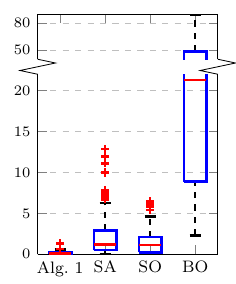}
  \caption{CPU time to obtain a function value within 1\% of the function value returned after final iteration.}
  \label{fig:tandem_cpu-a}
  \end{subfigure}
    \caption*{{\bf Model I}}

  \end{subfigure}
  	\setcounter{subfigure}{3}
  \begin{subfigure}[b]{0.99\linewidth}
 \begin{subfigure}[b]{.33\linewidth}
 \centering
  \includegraphics{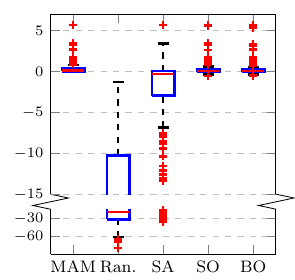}
  \caption{Difference between function value returned and value returned by Alg.~1. \\}
  \label{fig:tandem_obj-b}
  \end{subfigure}
      \hspace{2mm}
 \begin{subfigure}[b]{.33\linewidth}
 \centering
  \includegraphics{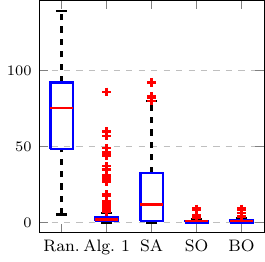}
  \caption{Euclidean distance between solution returned and true optimal solution.\\}
  \label{fig:tandem_sol-b}
  \end{subfigure}
      \hspace{2mm}
   \begin{subfigure}[b]{.33\linewidth}
 \centering
  \includegraphics{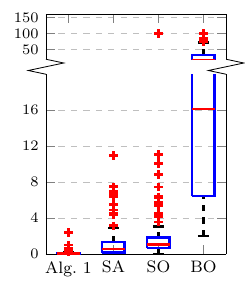}
  \caption{CPU time to obtain a function value within 1\% of the function value returned after final iteration.}
  \label{fig:tandem_cpu-b}
  \end{subfigure}
  \caption*{{\bf Model II}}
  \end{subfigure}
  \caption{\small{{\em(Tandem experiment 2)} Accuracy and efficiency of Alg.~1 and benchmarks for Model~I (top row) and Model~II (bottom row) for tandem networks.
Each box plot contains results for a specific method
{and}
displays results spanning 20 scenarios.} }
   \label{fig:tandem_results}
\end{figure}

Focusing first on accuracy, panels (a) and (b) of Figure~\ref{fig:tandem_results} display the differences between the function values returned by each method and the values returned by Algorithm~1 (including the function values corresponding to the random initial conditions). 
Values greater (less) than 0 exhibit better (worse) performance compared to Algorithm~1.
For these experiments, the reported objective function values are exact evaluations of the solutions outputted by the algorithms (using the methodology in Appendix~\ref{app:MAM}). 
The values given by the MAM method are the true optimal values. 
Since the experiment corresponding to each scenario is replicated 10 times with different initial conditions, each box plot summarises 200 data points. 
{Our key observations for the results in these panels are as follows.}
\begin{itemize}
\item[-] For Model~I, Algorithm~1 improved upon the the objective function values associated with the random initial conditions by a median amount of $13.2222$, an improvement which is approximately (within 1.5 units) matched by the other methods (and exactly equal for SA). 
This improvement falls short of the best possible improvement in the median (given by MAM as $14.7344$) by $1.5122$ units. 
For SO and BO this shortfall is smaller, at only $0.0527$ and $0.5357$, respectively. 
This means that the values returned by MAM, SO and BO are better than the values returned by Algorithm~1 in more than $50\%$ of the experiments (for this specific simple instance of Model~I), sometimes by up to 6.5855 units for SO and $4.3250$ units for BO (excluding outliers). 
\item[-] These differences in values (panel (a)) are reflected by the Euclidean distances from the optimal solution, as displayed in panel (b). 
All of the algorithms move the solution much closer to the optimal solution, relative to the random initial condition,
with median distances from the optimal solution of $3.8643$, $15.0830$, $1$ and $1.4142$ for Algorithm~1, SA, SO and BO, respectively. 
For Algorithm~1, SO and BO, these distances are also highly concentrated with interquartile ranges of $8.7097$, $2.2361$ and $3.8028$, respectively. 
The distances from the optimal solution are more spread out for SA, with an interquartile range of $39.7744$. 
\item[-] For Model~II, Algorithm~1 improved upon the the objective function values associated with the random initial conditions by a median amount of $19.6475$, an improvement which is approximately (within 0.3 units) matched by the other methods. 
This is close to optimal, lying within $0.1158$ units from the maximal median improvement (as given by MAM). 
Furthermore, since the interquartile ranges of the difference between MAM, SO, and BO with Algorithm~1 are all less than $0.35$, such results are achieved with very litte variability for these algorithms. 
The interquartile range of the difference in objective values is higher for SA at $3.0077$. 
\item[-] These differences in values (panel (d)) are again reflected in the Euclidean distances from the optimal solution, as displayed in panel (e). 
The median distance from the optimal solution for the random initial conditions of $75.4307$ is reduced down to $2$ by Algorithm~1 and 1 for SO and BO. 
The reduction is not as great for SA, achieving a median of difference of $11.6110$. 
These reductions are also achieved reliably for Algorithm~1, SO and BO, with interquartile ranges of $2.2839$, $1$ and $1.2071$, respectively. 
For SA, the reductions are less reliably achieved with an interquartile range of $31.6012$. 
\end{itemize}

We follow up on our observations of the solution accuracy from the four algorithms with an investigation into the computing power needed to obtain these solutions, where
panels (c) and (f) of Figure~\ref{fig:tandem_results} report the CPU times needed to achieve values that are within 1\% of the values reported in panels (a) and (d) of the figure. 
The reason we allow a 1\% gap is that SA, SO and BO may spend large amounts of CPU time making minor improvements in the objective value. 
Allowing this gap ensures that these algorithms are not penalised for such minimal refinements, thus providing a favourable case for these algorithms. 
{Our key observations for the results in these panels are as follows.}
\begin{itemize}
\item[-] Algorithm~1 is significantly  more efficient than the benchmarks for both models.
The median CPU times (seconds) for Algorithm~1 were $0.0978$ and $0.0783$ for Model~1 and Model~2, respectively.
This is much more efficient than the corresponding times for SA and SO, which across both models were in the range $0.5756$ to $1.2012$, and vastly more efficient than BO,
which had median CPU times of $21.275$ and $16.1226$ for Model~1 and Model~2, respectively.  
\end{itemize}

Based on the observations from the results {for the simple network} in this subsection,
we remark that there is evidence our algorithm and the benchmarks are all capable of substantially improving the performance of networks of
SLSs. 
SA appears to be the weakest algorithm in terms of both accuracy and efficiency (although it may be improved with further tuning,
we do not consider this since the other approaches do not benefit from tuning and it is not always possible to provide a tuning which works well for all perturbations of the model parameters and initial conditions). 
There is strong evidence that SO and BO reliably output a solution which is very close to the optimal solution. 
Our algorithm also outputs a solution which is close to the optimal solution, and it manages to do so much more efficiently than SA, SO or BO. 
Indeed, there is evidence that Algorithm~1 is typically (in terms of median performance) $10$ times faster than SA and SO, and $100$ times faster than BO. 
At this stage of our experiments, the results thus far suggest that our algorithm should be used when one needs a near optimal solution in a very efficient manner.
When a greater computational budget is available, our algorithm can be used to provide a high-quality starting point for SO, BO or SA following~\cite{Dieker2017} 
with the combination of algorithms being run to achieve asymptotic optimality.
We next conduct further experiments to examine whether these findings continue to hold true for more complex networks.

\subsection{Criss-cross networks with Markov-modulated-Poisson-process arrivals}\label{sec:CrissCrossExperiments}
To extend our investigation of the performance of our algorithm and the benchmarks into a setting with many stations and multiple customer classes, we consider a criss-cross network as displayed in Figure~\ref{fig:top2}.
This linear network consists of ten stations
{$\mathcal L = [10]$}
and two classes where the customers of one class follow the path $\psi_1 = (1,2,\dots,10)$ and the customers of the other class follow the path $\psi_2=(10,9,\dots,1)$. 
Customers arrive to the network according to a Markov modulated Poisson process (MMPP) with three background states;
arrivals occur at a rate of $10$, $20$ or $30$ depending on the state of the background process.
The background process jumps between states at a constant rate, and upon jumping moves to one of the other two states with equal probability. 
Arriving customers are equally likely to be of class $1$ or class $2$. 
The service requirements at each station are exponentially distributed. 
Since no analytical results are known for this network, we must rely on our benchmarks to test the accuracy of Algorithm~1. 

{\em Criss-cross Experiment~1 (illustrative).} 
We again begin with an experiment that illustrates the trajectories of our algorithm and benchmarks in terms of accuracy and efficiency. 
Consider the system above where the background process jumps between states at unit rate and the service rate at each station is $1.2$. 
In the case of Model~I, the reward for successfully processing a class 1 customer is  $\REV_1=9.8628$ and the reward for successfully processing a class 2 customer is $\REV_2=8.7339$
(where these parameter values were chosen, prior to conducting the experiment, uniformly at random from $(0,20)$). 
The capacity costs were sampled uniformly at random from $(0.1,0.3)$, as given in Table~\ref{tab:cc1CapCost}. 
In the case of Model~II, rewards for successfully processing customers are uniformly sampled (per class, per station) from $(0,10)$ and the costs ${\boldsymbol \COST}$ of capacity are uniformly sampled (per station) from $(0.1,\,0.3)$, as reported in Table~\ref{tab:cc2ex1param}. 

\begin{table}[H]
  \centering
  \fontsize{9}{9}\selectfont
  \caption{Criss-cross network Model~I experiment 1 capacity costs. }
\label{tab:cc1CapCost}
\begin{tabular}{lcccccccccc}\toprule
$l$ & 1 & 2 & 3 & 4 & 5 & 6 & 7 & 8 & 9 & 10 \\
$\COST_l$ & 0.20195&0.15437&0.16738&0.14339&0.1553&0.16866&0.27243&0.13134&0.12818&0.25142\\\bottomrule
\end{tabular}

\end{table}
\begin{table}[H]
  \centering
  \fontsize{9}{9}\selectfont
  \caption{Criss-cross network Model~II experiment 1 capacity costs. }
\label{tab:cc2ex1param}
\begin{tabular}{lcccccccccc}\toprule
$l$ & 1 & 2 & 3 & 4 & 5 & 6 & 7 & 8 & 9 & 10 \\
$\COST_l$ & 0.29064&0.12747&0.21388&0.29513&0.20067&0.23353&0.10684&0.19122&0.13117&0.19521\\
$\REV_{1,l}$ & 9.8628&5.0975&3.3692&2.7648&8.6216&1.4089&7.3632&3.4109&2.171&1.2418\\
$\REV_{2,l}$ & 8.7339&2.7184&2.1695&3.4332&1.567&7.5708&3.5566&6.668&5.6143&3.1974\\\bottomrule
\end{tabular}
\end{table}

\begin{figure}[h]
\centering
  \begin{subfigure}[b]{0.2\linewidth}
  \centering
 \includegraphics{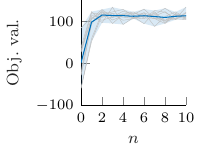}
  \caption{Alg.~1, Model~I.}
  \end{subfigure}
 \begin{subfigure}[b]{0.2\linewidth}
 \centering
 \includegraphics{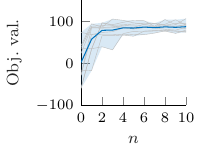}
  \caption{SA, Model~I.}
 \end{subfigure}
 \begin{subfigure}[b]{0.2\linewidth}
 \centering
 \includegraphics{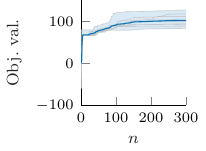}
  \caption{SO, Model~I.}
  \end{subfigure}
 \begin{subfigure}[b]{0.2\linewidth}
 \centering
 \includegraphics{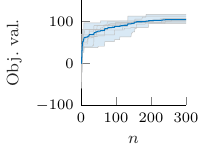}
  \caption{BO, Model~I.}
  \end{subfigure}

  \begin{subfigure}[b]{0.2\linewidth}
  \centering
 \includegraphics{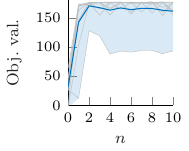}
  \caption{Alg.~1, Model~II.}
  \end{subfigure}
   \begin{subfigure}[b]{0.2\linewidth}
   \centering
 \includegraphics{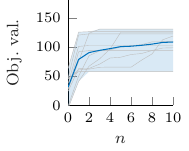}
  \caption{SA, Model~II.}
 \end{subfigure}
 \begin{subfigure}[b]{0.2\linewidth}
 \centering
 \includegraphics{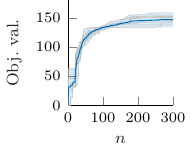}
  \caption{SO, Model~II.}
  \end{subfigure}
   \begin{subfigure}[b]{0.2\linewidth}
 \centering
 \includegraphics{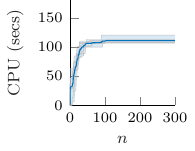}
  \caption{BO, Model~II.}
  \end{subfigure}
 \caption{Quality of solution outputted by Algorithm~\ref{ALG1final} and benchmarks for a criss-cross network (Criss-cross Experiment 1).}
 \label{fig:CrissCrossAcc}
\end{figure}

For each model of the networks just described, the initial capacity allocations were sampled uniformly at random from $(1,2,\dots,100)^{10}$. 
Each of these capacity allocations was used to initialise our algorithm and the benchmarks. 
In Figure~\ref{fig:CrissCrossAcc}, the improvement in the objective value from running the $3$ methods is displayed. 
Beyond explicitly displaying trajectories of the algorithms, the figure also displays the average of the sample paths and a shaded region containing all of the sample paths. 
{Our key observations for the results in this figure are as follows.}
\begin{itemize}
\item[-] {For Model~I, Algorithm~\ref{ALG1final}, SO and BO output comparable best objective values on average ($113.4697$, $102.4320$ and $104.4699$, respectively), which are substantially better than the average initial value (1.5387),
with Algorithm~1 demonstrating some advantage over the other two. 
The spread of the output values closely surround these average values (with the range of the final output values being $29.7253$, $32.3170$, $43.5180$ and $21.2739$ for Algorithm~1, SA, SO and BO, respectively).}
\item[-] For Model~II, Algorithm~1 {($161.5652$)} and SO {($147.4500$)} output somewhat similar best objective values on average, which are substantially better than SA {($108.6893$)}, BO {($111.0690$)} and the random initial {function value ($31.4809$)}.  Algorithm~1 again appears to have a sample path that has become trapped in a local solution. 
\item[-] {For both models}, SA improves the objective value relative to the random initial solution, but also appears to require many more iterations before it {may be} able to achieve the performance of our algorithm or SO
(which in turn requires vastly more CPU time than SO is allocated, as will be seen next). 
\end{itemize}

\begin{figure}[h]

  \begin{subfigure}[b]{0.2\linewidth}
   \centering
 \includegraphics{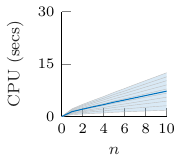}
  \caption{Alg.~1, Model~I.}
  \end{subfigure}
 \begin{subfigure}[b]{0.2\linewidth}
  \centering
 \includegraphics{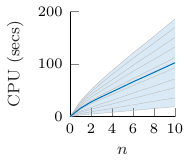}
  \caption{SA, Model~I.}
 \end{subfigure}
 \begin{subfigure}[b]{0.2\linewidth}
  \centering
 \includegraphics{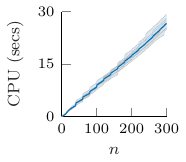}
  \caption{SO, Model~I.}
  \end{subfigure}
 \begin{subfigure}[b]{0.2\linewidth}
  \centering
 \includegraphics{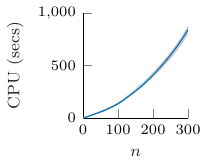}
  \caption{BO, Model~I.}
  \end{subfigure}

  \begin{subfigure}[b]{0.2\linewidth}
   \centering
 \includegraphics{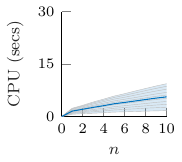}
  \caption{Alg.~1, Model~II.}
  \end{subfigure}
 \begin{subfigure}[b]{0.2\linewidth}
  \centering
 \includegraphics{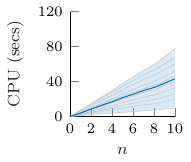}
  \caption{SA, Model~II.}
 \end{subfigure}
 \begin{subfigure}[b]{0.2\linewidth}
  \centering
 \includegraphics{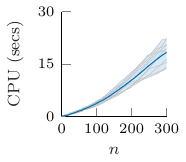}
  \caption{SO, Model~II.}
  \end{subfigure}
   \begin{subfigure}[b]{0.2\linewidth}
  \centering
 \includegraphics{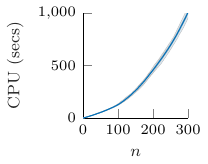}
  \caption{BO, Model~II.}
  \end{subfigure}

  \caption{Efficiency of Algorithm~\ref{ALG1final} and benchmarks for a criss-cross network (Criss-cross Experiment~1).}
 \label{fig:CrissCrossEff}
\end{figure}

We follow up on our above observations on the solution quality obtained by our algorithm and the benchmarks with an investigation into the computing power needed to obtain these solutions.
In Figure~\ref{fig:CrissCrossEff}, the cumulative CPU time needed to generate the objective values displayed in Figure~\ref{fig:CrissCrossAcc} are displayed for each sample path and on average.
{Our key observations for the results in this figure are as follows.}
\begin{itemize}
\item[-] BO uses vastly more CPU time and SA uses substantially more CPU time to generate iterates than does Algorithm~1 or SO (note the different scale of the axes). 
\item[-] For Model~I, the amount of CPU time needed by Algorithm~\ref{ALG1final} and SO is comparable. However, after an initial substantial improvement in value, SO endures a costly period of slow increase towards a best solution that is comparable to that given by Algorithm~\ref{ALG1final}. 
\item[-] Similarly, for Model~II, Algorithm~\ref{ALG1final} very quickly {(in terms of iterations and CPU usage)} reaches a best solution comparable to the best solution of SO {(as discussed above)}.
{SO  experiences a {slower} steady increase towards its best solution, requiring substantial CPU time.} 
\end{itemize}

This set of experiments has illustrated that our algorithm can more efficiently obtain solutions comparable {(and for this network topology they appear to be slightly better)} to  our benchmarks. 
Further evidence has been provided that SO is the superior benchmark. 
The potential for Algorithm~1 to become trapped in a local solution has appeared again. 
We now investigate whether these findings hold for a broader range of parameter values. 

{\em Criss-cross Experiment~2 (scenarios).} 
For each model we sampled 20 scenarios. 
In each scenario for both models,
the arrival rates corresponding to the three background states are sampled uniformly from $(0,20)$, $(0,40)$, and $(0,60)$. Rewards are sampled from $(0,10$), service rates (per station) from $(0,2.4)$, capacity costs from $(0.1,0.3)$, and the jump rate of the background chain from $(0,2)$, all uniformly at random. 

\begin{figure}[h!]
\centering
    	\setcounter{subfigure}{0}
\begin{subfigure}[b]{0.99\linewidth}
 \begin{subfigure}[b]{.33\linewidth}
 \centering
  \includegraphics{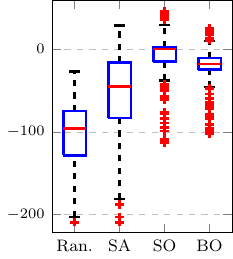}
  \caption{Difference between function value returned and value returned by Alg.~1. \\ }
  \label{fig:cc_obj-a}
  \end{subfigure}
  \hspace{2mm}
 \begin{subfigure}[b]{.33\linewidth}
 \centering
  \includegraphics{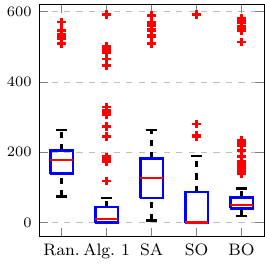}
  \caption{Euclidean distance between solution returned and the solution associated with the best observed function value.}
  \label{fig:cc_sol-a}
  \end{subfigure}
  \hspace{2mm}
   \begin{subfigure}[b]{.33\linewidth}
 \centering
  \includegraphics{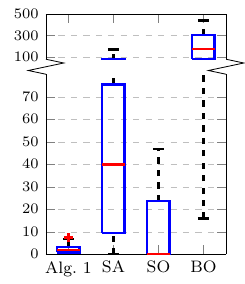}
  \caption{CPU time to obtain a function value within 1\% of the function value returned after final iteration.}
  \label{fig:cc_cpu-a}
  \end{subfigure}
    \caption*{{\bf Model I}}
  \label{fig:cc_results-s}
  \end{subfigure}
  	\setcounter{subfigure}{3}

  \begin{subfigure}[b]{0.99\linewidth}
 \begin{subfigure}[b]{.33\linewidth}
 \centering
  \includegraphics{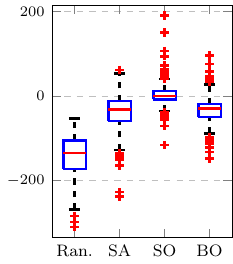}
  \caption{Difference between function value returned and value returned by Alg.~1. \\}
  \label{fig:cc_obj-b}
  \end{subfigure}
      \hspace{2mm}
 \begin{subfigure}[b]{.33\linewidth}
 \centering
  \includegraphics{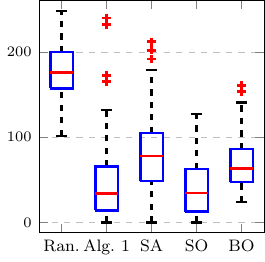}
  \caption{Euclidean distance between solution returned and the solution associated with the best observed function value.}
  \label{fig:cc_sol-b}
  \end{subfigure}
  \hspace{2mm}
   \begin{subfigure}[b]{.33\linewidth}
 \centering
  \includegraphics{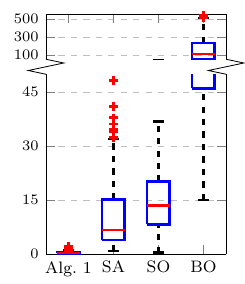}
  \caption{CPU time to obtain a function value within 1\% of the function value returned after final iteration.}
  \label{fig:cc_cpu-b}
  \end{subfigure}
  \caption*{{\bf Model II}}
  
  \end{subfigure}
  \caption{\small{{\em(Criss-cross experiment 2)} Accuracy and efficiency of Algorithm~1 and benchmarks for Model~I (top row) and Model~II (bottom row) for criss-cross networks. Each box plot contains results for a specific method.  Each box plot displays results spanning 20 scenarios. The experiment corresponding to each scenario is replicated 10 times with different initial conditions (which are held constant across methods), so that each box plot summarises 200 data points. } }
  \label{fig:cc_results}
\end{figure}

To investigate the accuracy of our algorithms, {as before, panels (a) and (d) of Figure~\ref{fig:cc_results} report the differences between the objective function values returned by each method and the values returned by Algorithm~1.  
All algorithms again improve substantially upon the random initial solutions. 
In contrast to the experiments for tandem networks however (as displayed in Figure~\ref{fig:tandem_results}), for both models in this case
Algorithm~1 clearly demonstrates higher accuracy relative to SA and BO (which have medians that are $45.3431$ and $17.7394$ lower than the median of Algorithm~1). 
For both models, Algorithm~1 and SO have essentially the same median objective function values (for Model~I the difference is in fact 0 and for Model~II SO is lower by only $0.3598$). 
The superiority of Algorithm~1 and SO in terms of accuracy is also evident in panels (b) and (e),
where it is shown that they return solutions which are typically much closer to the solution associated with the best observed function value across all experiments for any particular scenario. 
Finally, panels (c) and (f) of Figure~\ref{fig:cc_results} report the CPU times needed to achieve values which are within 1\% of the values underlying panels (a) and (d). 
For both models, it can be seen that SA and BO are substantially less efficient than Algorithm~1 and SO. 
For Model~I, in many cases SO and Algorithm~1 appear to have comparable (near 0 CPU time) efficiency,
but there are a clear set of cases where SO is much less efficient (taking up to $46.8400$ second and having a $75$$th$ percentile of $23.6700$ seconds). 
For Model~II, it is clear that Algorithm~1 is more efficient, except possibly in a small handful of cases.}
The key general observations made for tandem networks continue to hold true for criss-cross networks. 
Specifically, our algorithm and SO return the best solutions, and our algorithm {typically} does so much more efficiently than SO. 
The SA and BO benchmarks certainly improve the objective value but are much less efficient than our approach or SO. 

\subsection{Ring networks with MMPP arrivals}\label{sec:RingExperiments}
In this subsection we consider a network consisting of a set of $10$ stations arranged in a ring, as displayed in Figure~\ref{fig:top3}. 
Associated with arrivals at each station $i$ is a route $\psi_i = ( i, i+1, \dots, 1, 2,\dots, i-1)$ utilising all other stations in the network sequentially. 
This means that the system consists of $1000$ types of job when the jobs are classified by the combination of their route and the station at which they are being processed. 
We consider a parameterisation of the model where parameter values are randomly chosen following the same scheme as for criss-cross networks in the previous subsection (e.g., rewards are sampled from $(0,10)$).  

\begin{figure}[h!]
\centering
    	\setcounter{subfigure}{0}
\begin{subfigure}[b]{0.99\linewidth}
 \begin{subfigure}[b]{.33\linewidth}
 \centering
  \includegraphics{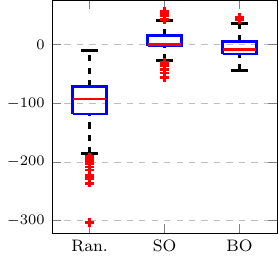}
  \caption{Difference between function value returned and value returned by Alg.~1. \\}
  \label{fig:ring_obj-a}
  \end{subfigure}
  \hspace{2mm}
 \begin{subfigure}[b]{.33\linewidth}
 \centering
  \includegraphics{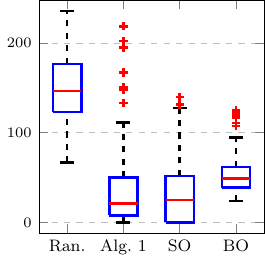}
  \caption{Euclidean distance between solution returned and the solution associated with the best observed function value.}
  \label{fig:ring_sol-a}
  \end{subfigure}
    \hspace{2mm}
   \begin{subfigure}[b]{.33\linewidth}
 \centering
  \includegraphics{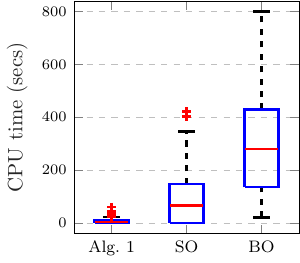}
  \caption{CPU time to obtain a function value within 1\% of the function value returned after final iteration.}
  \label{fig:ring_cpu-a}
  \end{subfigure}
    \caption*{{\bf Model I}}
  \label{fig:ring_results-s}
  \end{subfigure}
  	\setcounter{subfigure}{3}
\vspace{1mm}

  \begin{subfigure}[b]{0.99\linewidth}
 \begin{subfigure}[b]{.33\linewidth}
 \centering
  \includegraphics{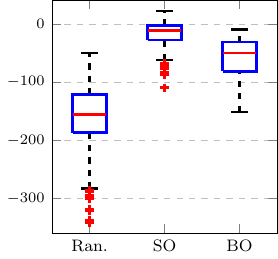}
  \caption{Difference between function value returned and value returned by Alg.~1. \\}
  \label{fig:ring_obj-b}
  \end{subfigure}
      \hspace{2mm}
 \begin{subfigure}[b]{.33\linewidth}
 \centering
  \includegraphics{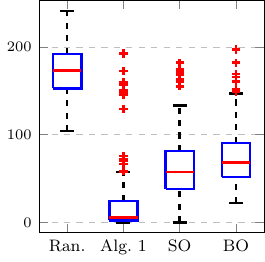}
  \caption{Euclidean distance between solution returned and the solution associated with the best observed function value.}
  \label{fig:ring_sol-b}
  \end{subfigure}
      \hspace{2mm}
   \begin{subfigure}[b]{.33\linewidth}
 \centering
  \includegraphics{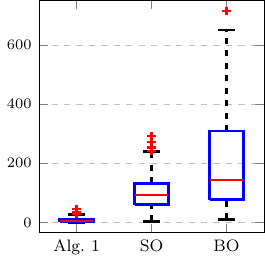}
  \caption{CPU time to obtain a function value within 1\% of the function value returned after final iteration.}
  \label{fig:ring_cpu-b}
  \end{subfigure}
  \caption*{{\bf Model II}}
  \end{subfigure}
  \caption{\small{{\em(Ring experiment)} Accuracy and efficiency of Algorithm~1 and benchmarks for Model~I (top row) and Model~II (bottom row) for ring networks. Each box plot contains results for a specific method.  Each box plot displays results spanning 20 scenarios. The experiment corresponding to each scenario is replicated 10 times with different initial conditions (which are held constant across methods), so that each box plot summarises 200 data points. } }
    \label{fig:ring_results}
\end{figure}

For the Model~I and Model~II variants of the ring network just described, $10$ initial capacity allocations were sampled uniformly at random from $(1,2,\dots,100)^{10}$ for each scenario.
Each of these $10$ capacity allocations was used to initialise our algorithm and the benchmarks for the ring network replications of the experiments previously {performed}
for the tandem (Figure~\ref{fig:tandem_results}) and criss-cross (Figure~\ref{fig:cc_results}) {networks}. 
Since we are optimising over a $10$ dimensional set, the results from the SA algorithm were not competitive in this case and therefore we only compare our algorithm to the SO and BO benchmarks. 
In Figure~\ref{fig:ring_results}, we see that {our} earlier findings {generalise to} these more complex systems;
{namely our algorithm returns comparable objective values to the best performing benchmark, which is SO. 
For Model~I, SO has a median performance that is only $0.3239$ greater than Algorithm~1, but it can also be up to $40.9869$ units better or up to $26.9052$ units worse. 
Similarly, for Model~II, SO has a median performance that is $11.0364$ units lower than
{Algorithm~1,}
but it can also be up to $22.6228$ units better or up to $109.5804$ units worse. 
For Model~I, the solutions returned by SO and Algorithm~1 have highly similar performance
(in panel (b) the medians and interquartile ranges for the distance to the best observed solution are within 3.7929 units and 9.2617 units of each other, respectively). 
In the case of Model~II, however, Algorithm~1 returns solutions that are much closer to the best observed solution (Algorithm~1: $5.3852$, SO: $57.4540$, BO: $68.0955$). 
Importantly, our algorithm achieves these results much more efficiently than SO and BO with median CPU usage for Model~I at $3.6807$, $65.9950$ and $279.0879$ for Algorithm~1, SO and BO, respectively;
and median CPU usage for Model~II at $5.4399$, $93.0550$ and $143.2500$ for Algorithm~1, SO and BO, respectively. 
The BO benchmark also returns good objective values, but these are noticeably inferior to those provided by Algorithm~1 or SO. 
In summary, the overwhelming superiority of Algorithm~1 in terms of efficiency is again evident---our approach is approximately at least $10$ times faster than SO or BO.
At the same time, Algorithm~1 returns excellent objective function values that are competitive with the benchmarks for Model~I and are usually superior to the benchmarks for Model~II. 
}

\subsection{Canonical networks with renewal arrivals and two-state Coxian distributed service requirements}\label{sec:CanonicalExperiments}
In this subsection we evaluate the performance of Algorithm~\ref{ALG1final} on a six-station, twelve-class system, as displayed in Figure~\ref{fig:top3}. 
Here $\boldsymbol c = (c_1, \dots, c_6)$,
{$\mathcal L = [6]$,}
{$\mathcal R = [12]$,}
and the customer classes follow
the network paths $\psi_1 =(1)$, $\psi_2=(1,3)$, $\psi_3=(1,3,5)$, $\psi_4 = (1,4)$, $\psi_5 = (1,4,5)$, $\psi_6=(1,4,6)$, $\psi_7 = (2)$, $\psi_8 = (2,4)$, $\psi_9=(2,4,6)$,
$\psi_{10} = (2,3)$, $\psi_{11} = (2,3,6)$, and $\psi_{12} = (2,3,5)$.
Customers of each class arrive to the first per-path station according to a per-class renewal process{;
the service time at each station is the same for all classes, while the service times differ across the various stations.}

In all of the experiments, the interarrival times and service requirements are generated from two-stage Coxian distributions with parameters set and scaled to match a collection of coefficients of variation (CoVs).
We consider $20$ randomly chosen (unique) scenarios where the interarrival times have CoVs randomly sampled from $\{0.75, 2, 3.25\}$, service times at stations $1$ and $2$ have CoVs randomly sampled from $\{2, 3.75, 5\}$,
service times at stations $3$ and $4$ have CoVs randomly sampled from $\{1.5, 3,4.5\}$, and service times at stations $5$ and $6$ have CoVs equal to $1.5$. 
The mean service requirements were randomly sampled (per station) from $(0,0.2)$.
The interarrival times for customers of class $i$ are sampled as $\frac{1}{2\lambda_i}\big(\,E_1+\frac{1}{q_i}E_2\,B_i\big)$, where $E_1$ and $E_2$ are independent unit-mean exponentially distributed
r.v.s, $q_i\in(0,1)$ is a real number, and $B_i$ is a Bernoulli r.v.\ with parameter $q_i$; the parameter $\lambda_i$ is sampled from $(0,0.2)$. 
Then, the parameter $q_i$ is set such that the CoV of the interarrival time matches the desired value for the scenario being studied.
The service times are specified on a per-station basis, with the service times of all customers at station $i$ sampled as $\frac{1}{2\mu_i}\big(\,E_1+\frac{1}{q_i}E_2\,B\big)$,
where $E_1$, $E_2$, $B_i$, and $q_i$ are as before (chosen such that the desired CoV is achieved).
The capacity costs $\COST_i$ are sampled from $(0.1,0.3)$ and the rewards ($\REV_r$, or $\REV_{r,i}$) are sampled from $(0,50)$. 

\begin{figure}[h!]
\centering
    	\setcounter{subfigure}{0}
\begin{subfigure}[b]{0.99\linewidth}
 \begin{subfigure}[b]{.33\linewidth}
 \centering
  \includegraphics{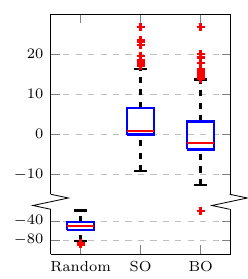}
  \caption{Difference between function value returned and value returned by Alg.~1. \\}
  \label{fig:canon_obj-a}
  \end{subfigure}
  \hspace{2mm}
 \begin{subfigure}[b]{.33\linewidth}
 \centering
  \includegraphics{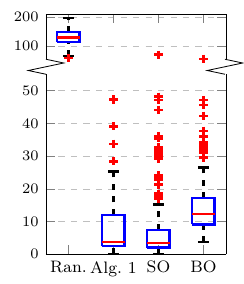}
  \caption{Euclidean distance between solution returned and the solution associated with the best observed function value.}
  \label{fig:canon_sol-a}
  \end{subfigure}
    \hspace{2mm}
   \begin{subfigure}[b]{.33\linewidth}
 \centering
  \includegraphics{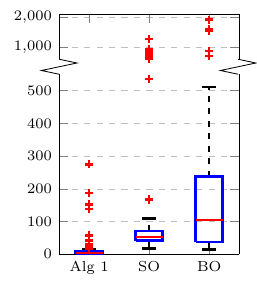}
  \caption{CPU time to obtain a function value within 1\% of the function value returned after final iteration.}
  \label{fig:canon_cpu-a}
  \end{subfigure}
    \caption*{{\bf Model I}}
  \label{fig:canon_results-a}
  \end{subfigure}
  	\setcounter{subfigure}{3}
\vspace{1mm}

  \begin{subfigure}[b]{0.99\linewidth}
 \begin{subfigure}[b]{.33\linewidth}
 \centering
  \includegraphics{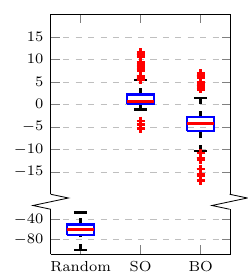}
  \caption{Difference between function value returned and value returned by Alg.~1. \\}
  \label{fig:canon_obj-b}
  \end{subfigure}
      \hspace{2mm}
 \begin{subfigure}[b]{.33\linewidth}
 \centering
  \includegraphics{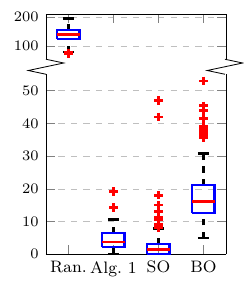}
  \caption{Euclidean distance between solution returned and the solution associated with the best observed function value.}
  \label{fig:canon_sol-b}
  \end{subfigure}
      \hspace{2mm}
   \begin{subfigure}[b]{.33\linewidth}
 \centering
  \includegraphics{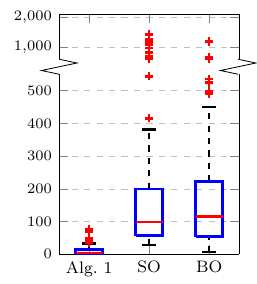}
  \caption{CPU time to obtain a function value within 1\% of the function value returned after final iteration.}
  \label{fig:canon_cpu-b}
  \end{subfigure}
  \caption*{{\bf Model II}}
  \label{fig:canon_results-b}
  \end{subfigure}
  \caption{\small{{\em(Canonical experiment)} Accuracy and efficiency of Algorithm~1 and benchmarks for Model~I (top row) and Model~II (bottom row) for canonical networks. Each box plot contains results for a specific method.  Each box plot displays results spanning 20 scenarios. The experiment corresponding to each scenario is replicated 10 times with different initial conditions (which are held constant across methods), so that each box plot summarises 200 data points. } }
    \label{fig:canon_results}
\end{figure}

For the canonical network just described, an initial capacity allocation was sampled from $(1,2,\dots,100)^6$ for each scenario. 
These capacity allocations were used to initialise trajectories of our algorithm and
{the SO and BO benchmarks}
for both model variants. 
The results from the SA algorithm were again not competitive; hence we only compare our algorithm to the SO and BO benchmarks.  
We performed similar experiments to those which resulted in Figure~\ref{fig:tandem_results}, Figure~\ref{fig:cc_results}, and Figure~\ref{fig:ring_results}. 
The results from these experiments, now in the context of the canonical networks of this section, are displayed in Figure~\ref{fig:canon_results}. 
{As with the other network types, we see for both models that our algorithm is able to find its best approximate solution much faster than SO and BO. 
For Model~I, we have that Algorithm~1, SO and BO respectively exhibit median CPU times (displayed in panel (c)) of $3.6967$, $52.5600$ and $104.0930$;
and for Model~II, we have that Algorithm~1, SO and BO respectively exhibit median CPU times (displayed in panel (f)) of $3.7164$, $98.3150$ and $114.8809$.
{All three algorithms}
obtain massive value improvements over the initial random capacity allocation (median more than $8.2910$ for Model~I and median more than $56.3184$ for Model~II).  
As shown in panels (a) and (d), the objective function values obtained by our algorithm are very similar to those obtained by SO
(median $0.7953$ units lower for Model~I, and median $0.6316$ units lower for Model~II) and typically slightly better than those obtained by BO
(median $2.1895$ units higher for Model~I, and median $4.3081$ units higher for Model~II). 
}
Algorithm~1 again demonstrates that it is able to obtain high-quality solutions much more efficiently than the benchmark approaches. 

\section{Conclusion}\label{sec:Conclude}
{In this paper we introduced a general class of networks of
SLSs,
motivated by applications arising in a wide variety of areas including cellular wireless networks, cloud edge computing, computer and network spare-part supply chains, and emergency services---just to name a few.
Since revenues are lost due to congestion and loss of customers and since costs are incurred due to allocating servers to network stations,
determining the optimal allocation is critically important and requires a careful balance between network performance and service costs. 
We therefore investigated the optimization of resource allocation in networks of
SLSs
with the goal of balancing the trade-off between mitigating congestion by increasing service capacity and maintaining low costs for the service capacity provided.
Two model variants of such networks are considered as being of particular interest.
The first model is applicable to situations where customers traverse a sequence of stations and may be lost at any one of them, e.g., cellular wireless networks or cloud edge computing,
whereas the second model is applicable to situations where customers are redirected to new stations until they find one with sufficient capacity, e.g., computer spare-part supply chains or emergency services.}

{Given the lack of analytical results and the computational burden of simulation-based methods, we proposed a novel hybrid functional-form approach
for the optimal allocation of resources
SLSs
that combines the speed of an analytical approach with the accuracy of simulation-based optimisation.
Our
core iterative algorithm
replaces
the computationally expensive gradient estimation in simulation optimisation with a closed-form analytical approximation that is calibrated using a single simple simulation run. 
Extensive computational experiments on complex networks within the context of four relevant example network topologies---two-station Markovian networks, criss-cross networks, ring networks,
canonical networks---show
that our approach consistently and with low variability yields near-optimal solutions for both network model variants under these four network topologies,
with objective function values that are comparable (and sometimes even superior) to those obtained using stochastic approximation, surrogate optimisation and Bayesian optimisation methods
while requiring dramatically less computational effort.}
%
%
{Hence, our general solution approach has several advantages:
use at finer time scales to obtain near-optimal solutions, including situations that require dynamic reallocation of capacities over frequently time-varying periods
(e.g., emergency services in a widespread disaster);
use at coarser time scales as a high-quality initial starting point for a simulation-based optimization approach;
use to address very large networks (e.g., $1000$s of stations or more) for which simulation-based optimization approaches can be infeasible due to their computational costs.}

Our {fundamental} approach can be extended to more general network {settings (e.g., incorporating re-trialling customers, priority classes or rerouting mechanisms)}.
In order to
{do so},
new functional-form approximations need to be formulated and
{evaluated}.
As we already considered two different model types, our study provides a solid foundation for these extensions and demonstrates how our general framework can be translated between models. \\

\noindent{\bf Acknowledgements.}
  An earlier version of this work was completed while BP was supported by the Australian Research Council (ARC) through the ARC Centre of Excellence for the Mathematical and Statistical Frontiers (ACEMS) under grant number CE140100049, the Netherlands Organisation for Scientific Research (NWO) through the Gravitation project N{\sc etworks} under grant number 024.002.003, and an Australian Government Research Training Program (RTP) scholarship. This version was completed while BP was supported by NWO grant 639.033.413 and ARC grant DP200101281. This work was initiated while BP was visiting IBM Research in Yorktown Heights. 

\bibliographystyle{abbrv}
\bibliography{references}

\begin{thebibliography}{10}

\bibitem{agrawal1996channel}
P.~Agrawal, D.~K. Anvekar, and B.~Narendran.
\newblock Channel management policies for handovers in cellular networks.
\newblock {\em Bell Labs Technical Journal}, 1(2):97--110, 1996.

\bibitem{amaran2016simulation}
S.~Amaran, N.~V. Sahinidis, B.~Sharda, and S.~J. Bury.
\newblock Simulation optimization: a review of algorithms and applications.
\newblock {\em Annals of Operations Research}, 240(1):351--380, 2016.

\bibitem{asmussen2007stochastic}
S.~Asmussen and P.~W. Glynn.
\newblock {\em Stochastic simulation: algorithms and analysis}, volume~57.
\newblock Springer Science \& Business Media, 2007.

\bibitem{axsater1990modelling}
S.~Axs{\"a}ter.
\newblock Modelling emergency lateral transshipments in inventory systems.
\newblock {\em Management Science}, 36(11):1329--1338, 1990.

\bibitem{baskett1975open}
F.~Baskett, K.~M. Chandy, R.~R. Muntz, and F.~G. Palacios.
\newblock Open, closed, and mixed networks of queues with different classes of
  customers.
\newblock {\em Journal of the ACM}, 22(2):248--260, 1975.

\bibitem{borst2004dimensioning}
S.~C. Borst, A.~Mandelbaum, and M.~I. Reiman.
\newblock Dimensioning large call centers.
\newblock {\em Operations research}, 52(1):17--34, 2004.

\bibitem{brotcorne2003ambulance}
L.~Brotcorne, G.~Laporte, and F.~Semet.
\newblock Ambulance location and relocation models.
\newblock {\em European journal of operational research}, 147(3):451--463,
  2003.

\bibitem{Brown1975}
M.~Brown and H.~Solomon.
\newblock A second-order approximation for the variance of a renewal reward
  process.
\newblock {\em Stochastic Processes and their Applications}, 3:301--314, 1975.

\bibitem{bull2011convergence}
A.~D. Bull.
\newblock Convergence rates of efficient global optimization algorithms.
\newblock {\em Journal of Machine Learning Research}, 12(10), 2011.

\bibitem{byrd2016stochastic}
R.~H. Byrd, S.~L. Hansen, J.~Nocedal, and Y.~Singer.
\newblock A stochastic quasi-newton method for large-scale optimization.
\newblock {\em SIAM Journal on Optimization}, 26(2):1008--1031, 2016.

\bibitem{Chiera2005}
B.~A. Chiera, A.~E. Krzesinski, and P.~G. Taylor.
\newblock Some properties of the capacity value function.
\newblock {\em SIAM Journal on Applied Mathematics}, 65(4):1407--1419, 2005.

\bibitem{Chiera2002}
B.~A. Chiera and P.~G. Taylor.
\newblock What is a unit of capacity worth?
\newblock {\em Probability in the Engineering and Informational Sciences},
  16(4):513--522, 2002.

\bibitem{Dieker2017}
T.~B. Dieker, S.~Ghosh, and M.~S. Squillante.
\newblock Optimal resource capacity management for stochastic networks.
\newblock {\em Operations Research}, 65(1):221--241, 2017.

\bibitem{gutmann2001radial}
H.-M. Gutmann.
\newblock A radial basis function method for global optimization.
\newblock {\em Journal of global optimization}, 19(3):201--227, 2001.

\bibitem{harrison1992brownian}
J.~M. Harrison and R.~J. Williams.
\newblock Brownian models of feedforward queueing networks: Quasireversibility
  and product form solutions.
\newblock {\em The Annals of Applied Probability}, pages 263--293, 1992.

\bibitem{harrison2005method}
J.~M. Harrison and A.~Zeevi.
\newblock A method for staffing large call centers based on stochastic fluid
  models.
\newblock {\em Manufacturing \& Service Operations Management}, 7(1):20--36,
  2005.

\bibitem{Hassin2015}
R.~Hassin, Y.~Y. Shaki, and U.~Yovel.
\newblock Optimal service-capacity allocation in a loss system.
\newblock {\em Naval Research Logistics (NRL)}, 62(2):81--97, 2015.

\bibitem{bookHe2014}
Q.~He.
\newblock {\em Fundamentals of Matrix-Analytic Methods}.
\newblock Springer, New York, 2014.

\bibitem{henderson2006handbooks}
S.~G. Henderson and B.~L. Nelson.
\newblock {\em Handbooks in operations research and management science:
  simulation}, volume~13.
\newblock Elsevier, 2006.

\bibitem{Jia2016}
M.~Jia, W.~Liang, Z.~Xu, and M.~Huang.
\newblock Cloudlet load balancing in wireless metropolitan area networks.
\newblock In {\em Proceedings of IEEE International Conference on Computer
  Communications (INFOCOM)}, 2016.

\bibitem{jones2001taxonomy}
D.~R. Jones.
\newblock A taxonomy of global optimization methods based on response surfaces.
\newblock {\em Journal of global optimization}, 21(4):345--383, 2001.

\bibitem{kelly1991loss}
F.~P. Kelly.
\newblock Loss networks.
\newblock {\em The annals of applied probability}, 1(3):319--378, 1991.

\bibitem{kiefer1952stochastic}
J.~Kiefer, J.~Wolfowitz, et~al.
\newblock Stochastic estimation of the maximum of a regression function.
\newblock {\em The Annals of Mathematical Statistics}, 23(3):462--466, 1952.

\bibitem{bookKleinrock1964}
L.~Kleinrock.
\newblock {\em Communication nets: Stochastic message flow and delay}.
\newblock McGraw--Hill, 1964.

\bibitem{Kranenburg2009}
B.~Kranenburg and G.-J. Van~Houtum.
\newblock A new partial pooling structure for spare parts networks.
\newblock {\em European Journal of Operational Research}, 199(3):908--921,
  2009.

\bibitem{kroese2014statistical}
D.~Kroese and J.~C. Chan.
\newblock {\em Statistical modeling and computation}.
\newblock Springer, 2014.

\bibitem{kwan2010mobility}
R.~Kwan, R.~Arnott, R.~Paterson, R.~Trivisonno, and M.~Kubota.
\newblock On mobility load balancing for {LTE} systems.
\newblock In {\em Proceedings of IEEE Vehicular Technology Conference (VTC)},
  2010.

\bibitem{mao2017survey}
Y.~Mao, C.~You, J.~Zhang, K.~Huang, and K.~B. Letaief.
\newblock A survey on mobile edge computing: The communication perspective.
\newblock {\em IEEE Communications Surveys \& Tutorials}, 19(4):2322--2358,
  2017.

\bibitem{mockus2012bayesian}
J.~Mockus.
\newblock {\em Bayesian approach to global optimization: theory and
  applications}, volume~37.
\newblock Springer Science \& Business Media, 2012.

\bibitem{Narayana1992}
S.~Narayana and M.~F. Neuts.
\newblock The first two moment matrices of the counts for the {M}arkovian
  arrival process.
\newblock {\em Communications in statistics. Stochastic models}, 8(3):459--477,
  1992.

\bibitem{Pang2015}
Z.~Pang, L.~Sun, Z.~Wang, E.~Tian, and S.~Yang.
\newblock A survey of cloudlet based mobile computing.
\newblock In {\em Proceedings of International Conference on Cloud Computing
  and Big Data}, 2015.

\bibitem{Patch2015}
B.~Patch, Y.~Nazarathy, and T.~Taimre.
\newblock A correction term for the covariance of renewal-reward processes with
  multivariate rewards.
\newblock {\em Statistics \& Probability Letters}, 102:1--7, 2015.

\bibitem{Patch2018transient}
B.~Patch and T.~Taimre.
\newblock Transient provisioning and performance evaluation for cloud computing
  platforms: A capacity value approach.
\newblock {\em Performance Evaluation}, 118:48--62, 2018.

\bibitem{Rahimi-Ghahroodi2017}
S.~Rahimi-Ghahroodi, A.~Al~Hanbali, W.~Zijm, J.~van Ommeren, and
  A.~Sleptchenko.
\newblock Integrated planning of spare parts and service engineers with partial
  backlogging.
\newblock {\em OR Spectrum}, pages 1--38, 2017.

\bibitem{razavi2012review}
S.~Razavi, B.~A. Tolson, and D.~H. Burn.
\newblock Review of surrogate modeling in water resources.
\newblock {\em Water Resources Research}, 48(7), 2012.

\bibitem{Restrepo2009}
M.~Restrepo, S.~G. Henderson, and H.~Topaloglu.
\newblock Erlang loss models for the static deployment of ambulances.
\newblock {\em Health care management science}, 12(1):67, 2009.

\bibitem{robbins1985stochastic}
H.~Robbins and S.~Monro.
\newblock A stochastic approximation method.
\newblock In {\em Herbert Robbins Selected Papers}, pages 102--109. Springer,
  1985.

\bibitem{Sanders2016}
J.~Sanders, S.~C. Borst, and J.~S. van Leeuwaarden.
\newblock Online network optimization using product-form markov processes.
\newblock {\em IEEE Transactions on Automatic Control}, 61(7):1838--1853, 2016.

\bibitem{shen2003joint}
Z.-J.~M. Shen, C.~Coullard, and M.~S. Daskin.
\newblock A joint location-inventory model.
\newblock {\em Transportation science}, 37(1):40--55, 2003.

\bibitem{sidi1997new}
M.~Sidi and D.~Starobinski.
\newblock New call blocking versus handoff blocking in cellular networks.
\newblock {\em Wireless networks}, 3(1):15--27, 1997.

\bibitem{sonmez2017analytical}
E.~S{\"o}nmez, A.~Scheller-Wolf, and N.~Secomandi.
\newblock An analytical throughput approximation for closed fork/join networks.
\newblock {\em INFORMS Journal on Computing}, 29(2):251--267, 2017.

\bibitem{vandenBerg2017}
P.~L. Van~den Berg, G.~A. Legemaate, and R.~D. van~der Mei.
\newblock Increasing the responsiveness of firefighter services by relocating
  base stations in amsterdam.
\newblock {\em Interfaces}, 47(4):352--361, 2017.

\bibitem{van2016comparison}
P.~L. Van~den Berg, J.~T. van Essen, and E.~J. Harderwijk.
\newblock Comparison of static ambulance location models.
\newblock In {\em Proceedings of Logistics Operations Management (GOL)}. IEEE,
  2016.

\bibitem{wein1988capacity}
L.~M. Wein.
\newblock Capacity allocation in generalized jackson networks.
\newblock {\em Operations Research Letters}, 8(3):143--146, 1989.

\end{thebibliography}

\appendix

\section{Matrix derivations for Markovian tandem network}\label{app:MAM}
While evaluating the objective functions \eqref{eq:OBJ_1} and \eqref{eq:OBJ_2} in Section~\ref{sec:Model} for a single-station,
single-class Markovian system (i.e., an Erlang-B system) is quite straightforward on a modern computer, this is not the case for more complex
{networks of
SLSs.}
The power of the approach described in \cite{Patch2018transient}, based on matrix-analytic methods (MAMs),
{is that more general systems can be considered.
In this section we use these methods to explicitly compute the objective functions for the relatively simple example provided in Section~\ref{sec:Model},
so that it can be used to provide an exact evaluation of the accuracy of our general approach as well as provide comparisons with the various benchmark methods.
%
More specifically, we explain how to use matrix-analytic methods to obtain}
explicit expressions for the reward terms in \eqref{eq:TandemOBJ_1} and \eqref{eq:TandemOBJ_2}, i.e., for
\begin{equation*}
\kappa_1 := \lambda_1\,\REV_{1}\,(1-p_{1,1}(c_1))(1-p_{1,2}(\boldsymbol c))\,, \quad
\kappa_{1,1}:= \lambda_1\,\REV_{1,2}\,(1-p_{1,1}(\boldsymbol c))\,, \quad
\kappa_{1,2}:= \lambda_1\,\REV_{1,2}\,(1-p_{1,2}(\boldsymbol c))\,p_{1,1}(c_1)\, ,
\end{equation*}
where $\kappa_1$ is the reward term in \eqref{eq:TandemOBJ_1} and $\kappa_{1,1}$ and $\kappa_{1,2}$ are the reward terms in \eqref{eq:TandemOBJ_2}.  


\subsection{Model~I}
\begin{figure}
   \begin{subfigure}[b]{.45\linewidth}
    \includegraphics[width = 0.95\textwidth]{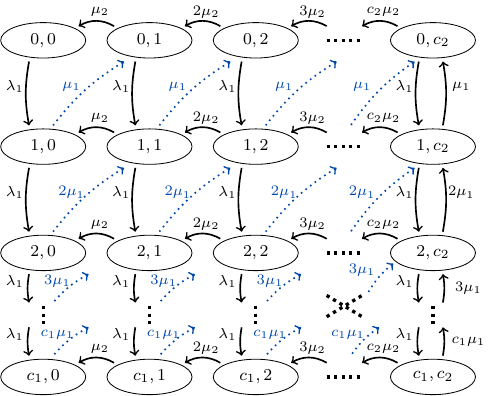}
    \vspace{3.5mm}
   \caption{Model I}
   \label{fig:mam_transitions_1}
   \end{subfigure}
  \begin{subfigure}[b]{.45\linewidth}
    \includegraphics[width = 0.95\textwidth]{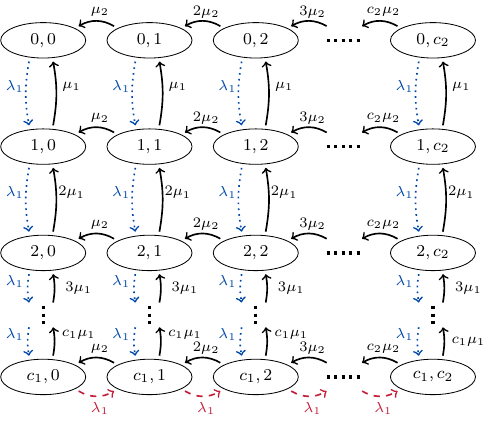}
    \caption{Model II}
   \label{fig:mam_transitions_2}
  \end{subfigure}
   \caption{Transition diagram for the bivariate Markov process $(X_1, X_2)$ underlying a Markovian tandem network of (a) Model~I and (b) Model~II. Blue (dotted) transitions result in rewards  $\REV_1$ (in (a)) or $\REV_{1,1}$ (in (b)), and red (dashed) transitions result in rewards $\REV_{1,2}$.}
  \label{fig:mam_transitions}
 \end{figure}

%
{First consider Model~I and let}
$\big(N_1(t),\,t\in\mathbb R_0\big)$ count the number of customers that successfully enter the second station. 
The rate at which $N_1$ increases corresponds to $\kappa_1$; this is the quantity we wish to identify. 
Let $X_{1}(t)$ and $X_{2}(t)$ be the number of customers that are processed by the system at the first and second stations, respectively. 
Upon an arrival to the first station, as long as $X_{1}<c_1$, then $X_{1}$ jumps up by $1$.
Upon a successful service completion at the first station, if $X_{2}<c_2$, then $X_{2}$ jumps up by $1$ and so too does $N_1$. 
It can be seen that the bivariate process $(X_{1}, X_{2})$ is a background process for an encompassing MAP that experiences arrivals when customers enter the second station (see, e.g., \cite[Section~2.3]{bookHe2014} for more details on MAPs). 
In order to find $\kappa_1$ explicitly, we need to carefully construct the structure of the Markov chain $(X_{1}, X_{2})$, use its relationship to $N_1$, and then apply
\cite[Theorem~1]{Narayana1992}. 
Figure~\ref{fig:mam_transitions_1} displays the transition diagram for the bivariate Markov chain $(X_1, X_2)$. 
The transitions in blue (dotted) correspond to service completions at the first station that result in a customer entering the second station, meaning $N_1$ jumps by $1$ for each occurrence of these transitions;
whereas the transitions of $(X_1,X_2)$ in black (solid) correspond to those that do not result in a jump in $N_1$.

In order to obtain an expression for $\kappa_1$, we require some additional notation. Let
\begin{itemize}
	\item $\boldsymbol 1_k$ be a $k$-tuple containing all unit entries;
	\item $I_k$ be a $k\times k$ identity matrix;
	\item $\bar {I}_k$ be a $k\times k$ matrix with upper diagonal containing unit entries and otherwise 0;
	\item $\underline {I}_k$ be a $k\times k$ matrix with lower diagonal containing unit entries and otherwise 0;
	\item $\kron$ denote the usual Kronecker product for matrices;
	\item $\text{o}(t) \to 0$ denote a function $h(t)$ such that $h(t)/t\to0$ as $t\to\infty$. 
\end{itemize}
Finally, using \cite[Theorem~1]{Narayana1992} (see also, e.g., \cite[Equation~(2.110)]{bookHe2014}), the expected number of customers successfully entering the second station is given by
\begin{align*}
\E N_1(t) &= {\boldsymbol \pi} \mathscr D_{1} \boldsymbol 1_{(c_1+1)(c_2+1)}\,t+\text{o}(t)\,,
\end{align*}
where $\boldsymbol \pi$ is the stationary distribution of the Markov chain generated by the infinitesimal generator $\mathscr D$.
Namely, $\boldsymbol \pi\,\mathscr D =0$ and $\boldsymbol \pi \boldsymbol 1_{(c_1+1)(c_2+1)} = 1$) with 
\[
\mathscr D = \mathscr D_0 + \mathscr D_1
\]
where 
\begin{equation*}
\mathscr D_0 = \text{diag}([\mathscr Q_0, \dots, \mathscr Q_{c_2}]) + \text{diag}([0,\dots,1])\kron \mathscr M_1 + I_{c_2+1} \kron \mathscr L + \mathscr M_2 \kron {I}_{c_1+1} \,,\qquad
\mathscr D_1 = \bar I_{c_2+1} \kron \mathscr M_1\,,
\end{equation*}
utilizing the $(c_1+1){\times}(c_1+1)$ matrices $\mathscr L$, $\mathscr M_1$, and $(\mathscr Q_j,\; j=0,\dots, c_2)$,
with non-zero entries $(\mathscr L)_{i,i+1} = \lambda_1$, $(\mathscr M_1)_{i+1,i} = i\mu_1$ and $(\mathscr Q_j)_{i,i}=-(\lambda_1+i\,\mu_1+j\,\mu_2)$ for
{$i \in [c_1]$,}
and utilizing the $(c_2+1){\times}(c_2+1)$ matrix $\mathscr M_2$ with non-zero entries $(\mathscr M_2)_{i+1,i} = i\mu_2$ for
{$i\in [c_2]$.}

Finally, the quantity of interest $\kappa_1$ follows immediately from differentiation of this expression and multiplication by the appropriate cost term $\COST_1$.
Based on this we can explicitly compute \eqref{eq:TandemOBJ_1} for a tandem network of loss systems of the Model~I variant.

\subsection{Model~II}
%
{Next consider Model~II and let}
$\big(N_1(t),\,t\in\mathbb R_0\big)$ and $\big(N_2(t),\,t\in\mathbb R_0\big)$ count the number of customers that successfully enter the first and second stations, respectively.
The rates at which $N_1$ and $N_2$ increase correspond to $\kappa_{1,1}$ and $\kappa_{1,2}$, respectively, which are the quantities we wish to identify. 
As above, let $X_{1}(t)$ and $X_{2}(t)$ be the number of customers that are processed by the system at the first and second stations, respectively. 
When an arrival occurs, as long as $X_{1}<c_1$, then $X_{1}$ jumps by 1 and, for this variant of the model, so does $N_1$. 
In the case that $X_1= c_1$ at the time of an arrival, if $X_{2}<c_2$, then $X_2$ and $N_2$ jump up by 1. 
Alternatively, if $X_1=c_1$ and $X_2=c_2$ at the time of an arrival, then $X_1$, $X_2$, $N_1$, and $N_2$ remain unchanged---the arrival is simply ignored. 
It can be seen in this case that the bivariate process $(X_{1}, X_{2})$ is a background process for the encompassing Marked MAP that experiences arrivals of different types
when customers enter the first or second station (see, e.g., \cite[Section~2.5]{bookHe2014} for more details on Marked MAPs). 
In order to explicitly find $\kappa_{1,1}$ and $\kappa_{1,2}$, we need to carefully devise the structure of the Markov chain $(X_{1}, X_{2})$, use its relationship to $N_1$ and $N_2$,
and then apply \cite[Theorem~1]{Narayana1992}. 
Figure~\ref{fig:mam_transitions_2} displays the transition diagram for the bivariate Markov chain $(X_1, X_2)$. 
The transitions in blue (dotted) correspond to accepted arrivals at the first station, meaning $N_1$ jumps by $1$ for each occurrence of these transitions;
the transitions in red (dashed) correspond to accepted arrivals at the second station, meaning $N_2$ jumps by $1$ for each occurrence of these transitions; and
the transitions of $(X_1,X_2)$ in black (solid) correspond to those which do not result in an increase of either $N_1$ or $N_2$.  

Then, using \cite[Theorem~1]{Narayana1992} (see also, e.g., \cite[Equation~(2.110)]{bookHe2014}), the expected number of customers successfully entering the first and second stations are given by
\begin{equation*}
\E N_1(t) = \tilde{\boldsymbol \pi} \tilde{\mathscr D}_{1} \boldsymbol 1_{(c_1+1)(c_2+1)}\,t+\text{o}(t)\,,\qquad
\E N_{2}(t) = \tilde{\boldsymbol \pi} \tilde{\mathscr D}_{2} \boldsymbol 1_{(c_1+1)(c_2+1)}\,t+\text{o}(t)\,,
\end{equation*}
respectively,
where $\tilde{\boldsymbol \pi}$ is the stationary distribution of the Markov chain generated by the infinitesimal generator $\tilde{\mathscr D}$.
Namely, $\tilde{\boldsymbol \pi}\,\tilde{\mathscr D} =0$ and $\tilde{\boldsymbol \pi} \boldsymbol 1_{(c_1+1)(c_2+1)} = 1$) with 
\[
\tilde{\mathscr D} = \tilde{\mathscr D}_0 + \tilde{\mathscr D}_1+ \tilde{\mathscr D}_2
\]
where 
\begin{equation*}
\tilde{\mathscr D}_0 = \text{diag}([\mathscr Q_0, \dots, \mathscr Q_{c_2}]) + \mathscr M_2\kron {I}_{c_1+1}+I_{c_2+1}\kron\mathscr M_1-\mathring I_{c_2+1}\kron\Lambda\,,\quad
\tilde{\mathscr D}_1 = I_{c_2+1}\kron \mathscr L\,,\quad
\tilde{\mathscr D}_2 = \bar I_{c_2+1} \kron \Lambda\,.
\end{equation*}
Here we utilize the matrices defined previously for Model~I with $\Lambda$ being a $(c_1+1){\times}(c_1+1)$ matrix of zeros except for $(\Lambda)_{c_1+1,c_1+1}=\lambda_1$,
and $\mathring I_{c_1+1}$ a $(c_2+1){\times}(c_2+1)$ matrix with unit diagonal entries except for $(\mathring I_{c_1+1})_{c_2+1,c_2+1}=0$. 


\section{Stochastic approximation implementation}\label{app:StochApprox}
For $\mathcal C \subset \mathbb R^d$, let $\Pi_{\mathcal C}$ be a function $\mathbb R^d \to \mathcal C$ that, for $\boldsymbol x\in \mathbb R^d$, returns $\boldsymbol c\in\mathcal C$ which minimises $||\boldsymbol c-\boldsymbol x||$.
Let $\mathbf e_l$ be a vector with all components $0$ except for component $l$, which is unitary, and
let $\{\delta^{(n)}\}_{n\in\Ints^+}$ and $\{\alpha^{(n)}\}_{n\in\Ints^+}$ be sequences satisfying
\begin{equation}\label{eq:ConvSAcond}
\sum_{n=1}^\infty \alpha^{(n)} = \infty\,, \qquad \sum_{n=1}^\infty {\alpha^{(n)}}{\delta^{(n)}} < \infty\,, \qquad \sum_{n=1}^\infty {\alpha^{(n)}}^2{\delta^{(n)}}^{-2} < \infty\,.
\end{equation}
Then, if $\hat f$ are uniformly bounded r.v.s,
Algorithm~\ref{ALG_SA} is a convergent (in probability) stochastic approximation algorithm \cite{kiefer1952stochastic}.

\begin{algorithm}
\SetAlgoLined
\KwResult{Approximation to optimal capacity allocation.}
Choose $\boldsymbol c^{(0)}\in (0,\infty)^L,\,\epsilon>0,\,N\in\Ints^+$\;
Set $n=1$ and $g > \epsilon$\;
 \While{$g >\epsilon$ and $n\le N$}{
 	Determine estimates
${(\widehat{\nabla_{\boldsymbol c} f}(\boldsymbol c^{(n-1)}))}_l = \frac{\hat f(\boldsymbol c^{(n-1)}+\delta^{(n)}\mathbf e_l)-\hat f(\boldsymbol c^{(n-1)}-\delta^{(n)}\mathbf e_l)}{2\delta^{(n)}}$
for components
{$l\in [L]$}
of the Jacobian\;
 	Choose a step size $\alpha^{(n)}$\;
Set
$\boldsymbol c^{(n)} = \Pi_\mathcal C(\boldsymbol c^{(n-1)} + \alpha^{(n)}\widehat{\nabla_{\boldsymbol c} f}(\boldsymbol c^{(n-1)}))$\;
Set $g = ||\boldsymbol c^{(n-1)}-\boldsymbol c^{(n)}||$ and $n=n+1$\;
 }
  Output $\boldsymbol c^{(n)}$.
 \caption{Stochastic approximation.}
 \label{ALG_SA}
\end{algorithm}

In our implementation for each iteration of this algorithm,
the step size $\alpha^{(n)}$ is determined using the following backtracking line-search method. First choose $\beta>0$, $\rho_1\in(0,1)$, $\rho_2\in(0,1)$, $D>0$. Set $\alpha^{(n)}=\beta^{(n)}$ and $d=1$, and then:
\begin{enumerate}
\item  Determine $F = \hat f\left(\Pi_\mathcal C\left(\boldsymbol c^{(n-1)}+\alpha^{(n)}\widehat{\nabla_{\boldsymbol c} f}(\boldsymbol c^{(n-1)}\right)\right)$;
\item  If $F \ge \hat f(\boldsymbol c^{(n-1)})+\rho_2\alpha^{(n)}\widehat{\nabla_{\boldsymbol c} f}(\boldsymbol c^{(n-1)})\widehat{\nabla_{\boldsymbol c} f}(\boldsymbol c^{(n-1)})^{\sf T}$ or $d \ge D$: output $\alpha^{(n)}$;\\
Else: set $\alpha^{(n)} = \rho_1\alpha^{(n)}$ and $d = d+1$, and return to {\sc (i)}.
\end{enumerate}
Choosing $\beta^{(n)} = \beta n^{-1/3}$ and $\delta^{(n)} =\delta n^{-1/6}$ where $\beta,\delta\in\mathbb R_+$ are positive real numbers ensures that \eqref{eq:ConvSAcond} holds and the algorithm therefore asymptotically converges to the true optimiser with probability 1.
In all of our experiments we used $\beta=150$, $\delta=5$, $\rho_1=0.8$, $\rho_2=0.5$ and $D=20$. 

\end{document}